\documentclass{aa}  

\usepackage{graphicx}
\usepackage{txfonts}
\usepackage{hyperref}
\begin{document} 

    \title{Connecting clustering and the cosmic web:} 
   \subtitle{Observational constraints on secondary halo bias} 


 \author{Facundo Rodriguez\thanks{facundo.rodriguez@unc.edu.ar}
   \inst{1,2}
          \and
          Antonio D. Montero-Dorta\thanks{amonterodorta@gmail.com}\inst{3}
          }

   \institute{CONICET. Instituto de Astronomía Teórica y Experimental (IATE). Laprida 854, Córdoba X5000BGR, Argentina.
   \and
             Universidad Nacional de Córdoba (UNC). Observatorio Astronómico de Córdoba (OAC). Laprida 854, Córdoba X5000BGR, Argentina.
        \and
        Departamento de F\'isica, Universidad T\'ecnica Federico Santa Mar\'ia, Avenida Vicu\~na Mackenna 3939, San Joaqu\'in, Santiago, Chile.
             }


  \abstract
    {Cosmological simulations predict significant secondary dependencies of halo clustering on a variety of internal halo properties, as well as on environmental factors. Detecting these often subtle signals in observational data remains a significant challenge, with important ramifications for both galaxy evolution and cosmology.}
   {We aim to probe secondary halo bias in observational survey data, employing galaxy groups as proxies for dark matter haloes. We quantify the strength of multiple secondary bias signals defined by the colour of the central galaxy and various environmental diagnostics.}
   {We employ an extended and refined galaxy group catalogue constructed from the Sloan Digital Sky Survey. Secondary bias is defined as any deviation in the clustering strength of groups at fixed mass, quantified through the projected two-point correlation function. Our environmental analysis uses the DisPerSE cosmic-web reconstruction algorithm to compute the distances to the critical points of the density field and incorporates local group overdensity measurements on multiple spatial scales.}
   {We robustly detect several forms of secondary bias in the clustering of galaxy groups. At fixed mass, groups hosting red central galaxies are more strongly clustered than those with blue centrals, with $b_{\rm relative}$ ranging from $\sim 1.2$ for the 15\% reddest centrals to $\sim 0.8$ for the bluest ones. Environmental dependencies based on cosmic--web distances are also present, though significantly weaker and largely mass-independent. The strongest signal arises from local overdensity: groups in the densest 15\% of environments reach $b_{\rm relative} \sim 1.4$, while those in the least dense regions fall to $b_{\rm relative} \sim 0.7$. These results establish a clear observational hierarchy for secondary halo bias.}
  {The colour of central galaxies correlates with the local group overdensity, which in turn correlates with the bias at fixed group mass. Assuming that central galaxy colour traces halo assembly history, this three-stage picture offers a conceptual link between our results and halo assembly bias.}


   \maketitle
%

\section{Introduction}

The large-scale linear bias of dark matter (DM) haloes encodes fundamental information on the process of halo formation from the underlying matter density field. Halo bias, which can be measured by comparing the clustering amplitude of haloes and the matter field on large scales, depends primarily on the peak height of density fluctuations, $\nu$, as it can be analytically derived from structure formation formalisms (see, e.g., \citealt{Kaiser1984,Bardeen1986,Mo1996, ShethTormen1999, Sheth2001, Tinker2010}). This fundamental dependence is naturally reflected on the halo virial mass: more massive haloes are more tightly clustered (have a higher bias) than less massive haloes. At fixed $\nu$ or halo mass, halo bias has been shown in cosmological simulations to depend on a variety of \emph{secondary} halo properties, including formation time, concentration, spin, or shape (see, e.g., \citealt{Sheth2004,gao2005,Wechsler2006,Gao2007,Dalal2008, Angulo2008,Li2008,faltenbacher2010, Lazeyras2017,2018Salcedo,Mao2018, Han2018,SatoPolito2019, Johnson2019, Paranjape2018, Ramakrishnan2019,Contreras2019, Tucci2021,  Contreras2021_cosmo, MonteroDorta2021, MonteroRodriguez2024, Balaguera2024}). The secondary dependence of bias on internal halo properties beyond halo mass extends back to the initial field, i.e., to the properties of the initial collapsing regions that give rise to haloes \citep{MonteroDorta2025}.

As can be also inferred from analytical models of the evolution of structure, halo bias is known to depend on the local and, naturally, large-scale distribution of matter. At fixed halo mass, the dependencies on the local density across multiple scales, on the geometry of the tidal field and its anisotropic (traceless) component, and on the specific location of haloes within the cosmic web have been extensively characterised in simulations (e.g., \citealt{Borzyszkowski2017, Musso2018, Paranjape2018, Ramakrishnan2019, MonteroRodriguez2024, Balaguera2024}). Recognizing that linear bias, overdensity, and cosmic-web geometry all reflect the same underlying density field, in this work we distinguish between local overdensity and cosmic-web architecture to isolate their specific impacts on secondary halo bias.
In \cite{MonteroRodriguez2024}, in particular, we used the IllustrisTNG hydrodynamical simulation to measure secondary bias based on the distance of central galaxies (i.e., distinct haloes) to the critical points of the density field, identified through a cosmic web reconstruction with the Discrete Persistent Structures Extractor \citep[DisPerSE;][]{Sousbie2011a,Sousbie2011b}. The resulting dependencies at fixed halo mass are shown to be very significant, largely exceeding the signal measured for \emph{halo assembly bias}, the secondary dependence of halo bias on the assembly history of haloes at fixed mass.

The complete set of internal and environmental dependencies of halo clustering at fixed halo mass, which we call here \emph{secondary halo bias}, provides a detailed description of the halo-matter connection, which has important applications for cosmology \citep{Wechsler2018}. Observationally, the dependence of galaxy clustering on galaxy properties and environment has been extensively investigated using spectroscopic surveys, showing that galaxies of higher mass, redder colours, early-type morphologies and those in denser regions are more tightly clustered across a wide range of scales than the rest of the population (e.g., \citealt{AbbasSheth2006,Meneux2006,Coil2006, Abbas2007, Skibba2009, Zehavi2011, Hartley2013, McNaughtRoberts2014, Zhai2023}). Specifically, \cite{Zehavi2011} used the SDSS to show that, at fixed luminosity, redder galaxies exhibit a higher-amplitude and steeper correlation function, a trend consistent with an increased satellite fraction in massive haloes. These general trends persist mostly when the analysis is performed at fixed stellar mass (e.g., \citealt{Li2006, Zehavi2011, LawSmith2017}). Serious challenges arise, however, when attempting to study these trends at fixed halo (or group) mass, largely due to the observational uncertainties involved in estimating halo masses. Potential dependencies of galaxy clustering at fixed halo mass that can be directly attributed to secondary halo properties (such as concentration or formation time) are often considered manifestations of \emph{galaxy assembly bias}, a phenomenon that remains to be robustly established observationally. 

The study of secondary bias at fixed halo mass using observations has yielded a variety of results, often depending on the selection method and the proxy used for halo mass. For example, \cite{Wang2008}, using an SDSS galaxy group catalogue, reported a dependence of group clustering on central galaxy colour at fixed group mass, with groups having red centrals clustering more strongly, particularly in less massive haloes. These results are in line with those of \cite{MonteroDorta2017}, who did report such a dependence, albeit for luminous red galaxies from the Baryon Oscillation Spectroscopic Survey (BOSS; \cite{Dawson2013}); a trend later shown by \cite{Niemiec2018} to be consistent with a pattern occurring at fixed halo mass based on weak lensing. Compatible results splitting the galaxy population using the stellar age were found by \cite{Lacerna2014}. In contrast, other observational studies have found no evidence for a significant secondary signal. Using the Sloan Digital Sky Survey (SDSS; \citealt{York2000}) and the \cite{Yang2007} group catalogue, \cite{Lin2016} found no evidence for a dependence of galaxy clustering on star formation history, for samples selected at approximately fixed halo mass using weak lensing. More recently, \citet{Sunayama2022} also reported no significant secondary signals when probing clustering dependencies at fixed halo mass using a variety of central galaxy properties from the redMaPPer cluster catalogue \citep{Rykoff2014} from the SDSS.

In the context of galaxy assembly bias, secondary dependencies at fixed halo mass have been reported, using indirect methods, in the anisotropic clustering of massive galaxies, \cite{Obuljen2019,Obuljen2020}, halo occupation statistics (also known as occupancy variations\footnote{The concept and specific term occupancy variation were first introduced and defined in the studies by \citep{Zehavi2018} and \cite{Artale2018}.}, \citealt{Yuan2021,Wang2022,Pearl2024}), and the stellar-to-halo mass relation \citep{Oyarzun2024}. Focusing on halo spin and applying a novel proxy based on the coherent motion of galaxies inside and around clusters, \citet{Kim2025} also presented observational indications of halo spin bias, the secondary dependence of halo bias on spin at fixed halo mass. 

In this paper, we use an extended SDSS group catalogue based on a refined version of the \citet{rodriguez2020} method to provide observational constraints on secondary halo bias, understood here as any potential dependence of group (central galaxy) bias at fixed group mass. A key challenge stems from the uncertainty in the estimation of halo masses, an inherent limitation affecting many observational studies of galaxy and halo bias. This includes analyses that seek to detect assembly bias from galaxy clustering, especially those that do not explicitly rely on mock catalogues for validation. In our work, we employ halo masses derived from abundance matching. Although not a direct measurement of mass, the mass calibration in our group catalogue has been shown to align well with weak gravitational lensing estimates \citep[e.g.][]{Gonzalez2021} and is widely considered an effective indicator of the mass of the halo hosting the central galaxy.

More restrictive definitions of secondary bias have been adopted in the literature. However, as mentioned above, we adopt a broad and operational definition of secondary halo bias that encompasses both environmental and central galaxy properties. This definition is intentionally generic and primarily motivated by observational considerations, allowing us to compare different dependencies within the same framework consistently. We explore several secondary dependencies, including both central galaxy properties and environmental metrics. To characterise the environment, we use local density measurements on multiple physical scales, as well as the distance of galaxies to the critical points of the density field, as defined by the cosmic web reconstruction from DisPerSE \citep{Malavasi2020a}. Our work provides key observational constraints on secondary bias, establishing a hierarchy of dependency strengths that can be directly compared with simulations. Previous studies have, in fact, explored this environmental dependence with different approaches. In simulations, \cite{Paranjape2018} proposed the key role of tidal anisotropy as a mediator between bias and internal halo properties, whereas \cite{MonteroRodriguez2024} provided detailed predictions for the dependence of clustering on the distance to the DisPerSE critical points in the IllustrisTNG hydrodynamical simulation. On the observational side, \citet{alam2019} measured the dependence of clustering on several of these environmental properties and used these results to constrain modified halo occupation distribution (HOD) models aimed at testing assembly bias. Our analysis complements these efforts by providing a direct measurement, using a cluster catalogue to achieve precise halo mass control and isolate the geometric influence of the density and cosmic web.

This paper is organized as follows. In Section~\ref{sec:data}, we describe the data used, including the SDSS DR18 galaxy catalogue and our extended group catalogue, as well as the cosmic web information from DisPerSE. Section~\ref{sec:secondarybias} details the methodology followed to measure secondary bias based on the ratios of the correlation functions. Section~\ref{sec:results} presents our main results, focusing on the dependence of secondary bias on intrinsic (colour) and environmental properties (distances to critical points and local density), accompanied by a robustness analysis. Finally, in Section~\ref{sec:conclusions}, we discuss the implications of our findings and present our conclusions. Throughout this work, we adopt the standard $\Lambda$CDM cosmology \citep{Planck2016}, with parameters $\Omega_m = 0.3089$, $\Omega_b = 0.0486$, $\Omega_\Lambda = 0.6911$, $H_0 = 100\, h \, \rm km \, \rm s^{-1} \, \rm Mpc^{-1}$ where $h = 0.6774$, $\sigma_8 = 0.8159$ and $n_s = 0.9667$.

\begin{figure}
    \centering
    \includegraphics[width=0.9\columnwidth]{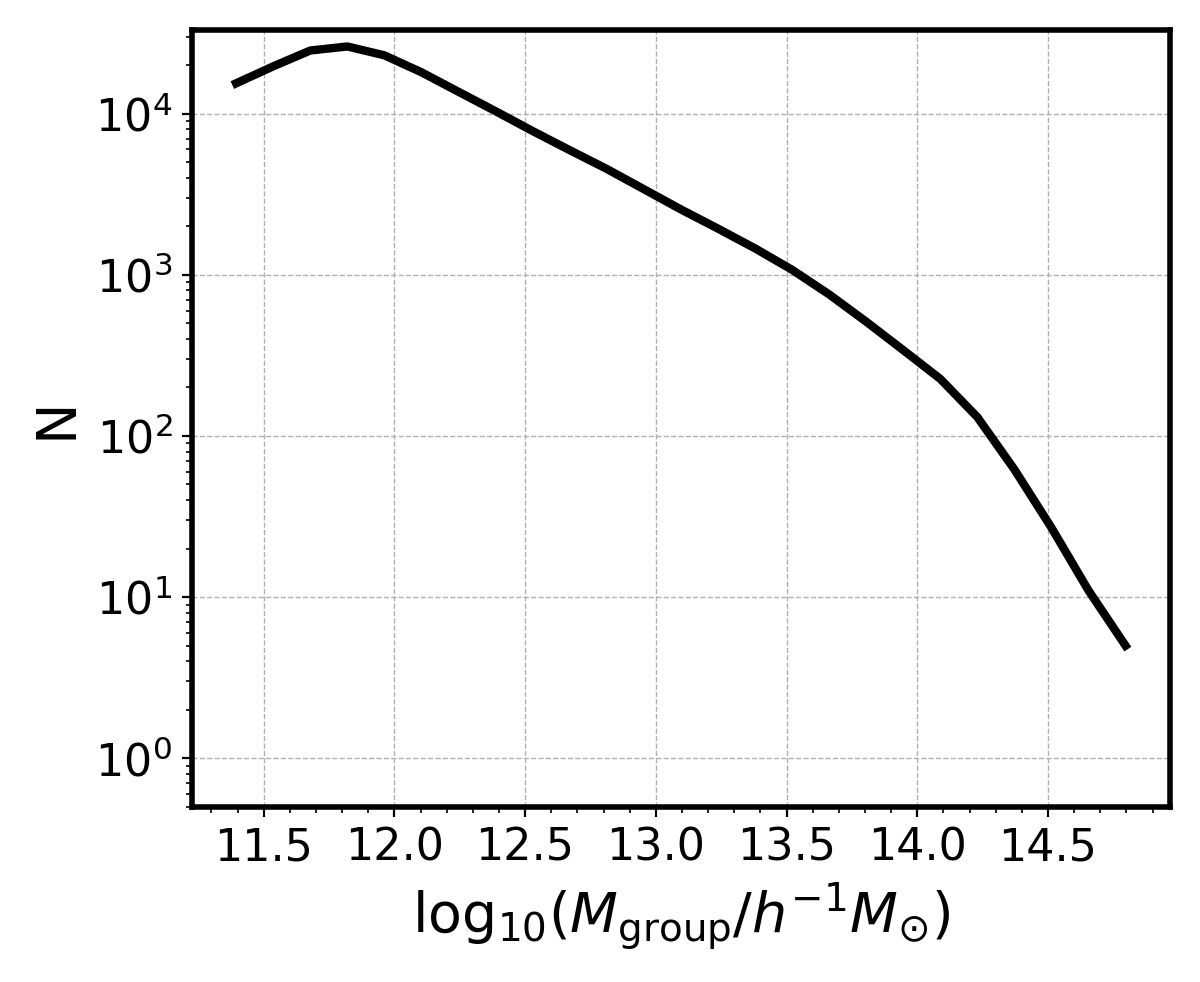}
    \includegraphics[width=0.9\columnwidth]{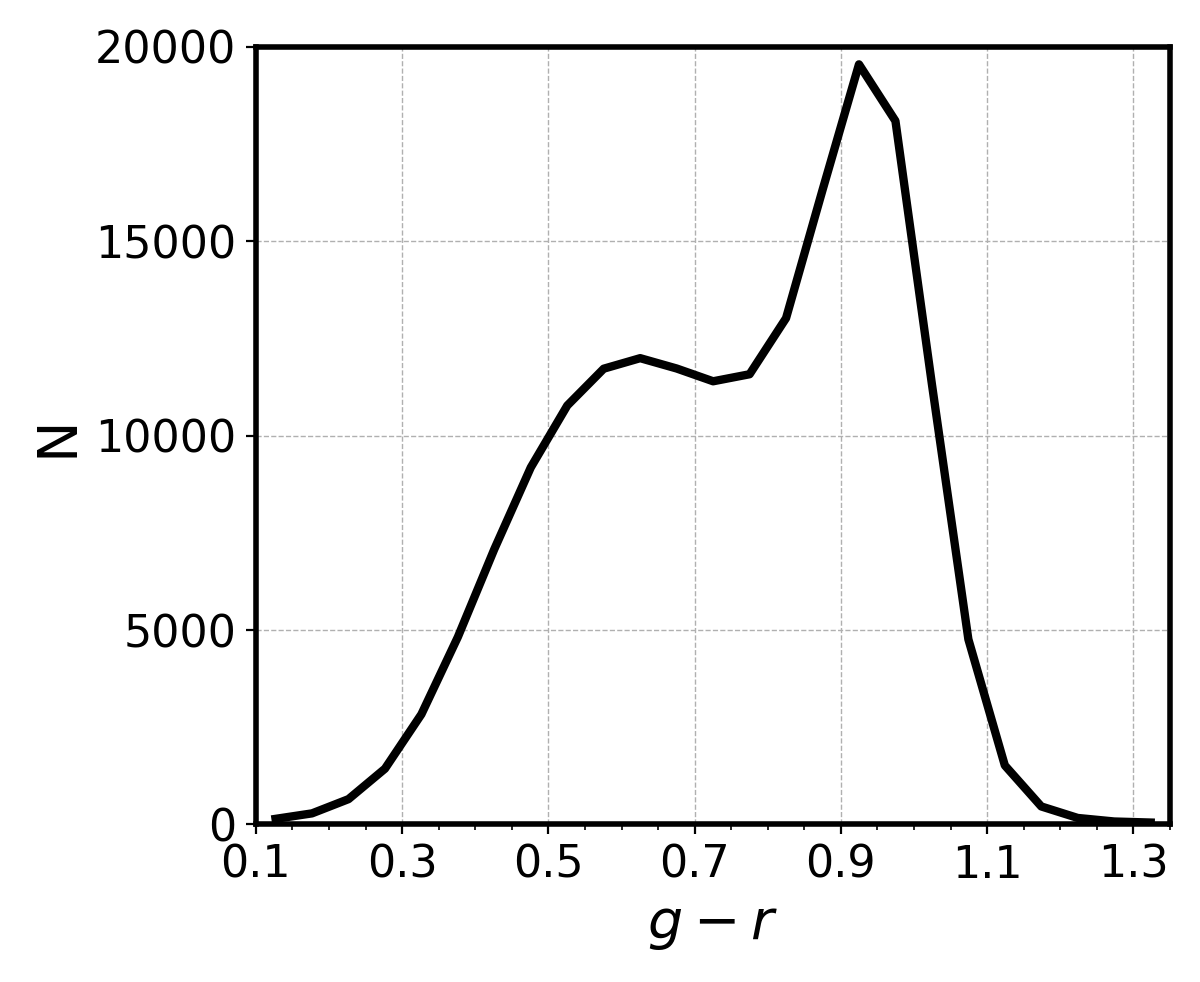}
    \caption{Main properties of the galaxy groups analysed in this study. The top panel presents the group mass distribution, which exhibits the characteristic decreasing shape. The bottom panel displays the colour ($g-r$) distribution of the central galaxies within these groups. The bimodal nature of this distribution clearly distinguishes between the red and blue galaxy subpopulations.
}
    
    \label{Grupdists}%
\end{figure}

 \begin{figure}
    \centering
    \includegraphics[width=0.9\columnwidth]{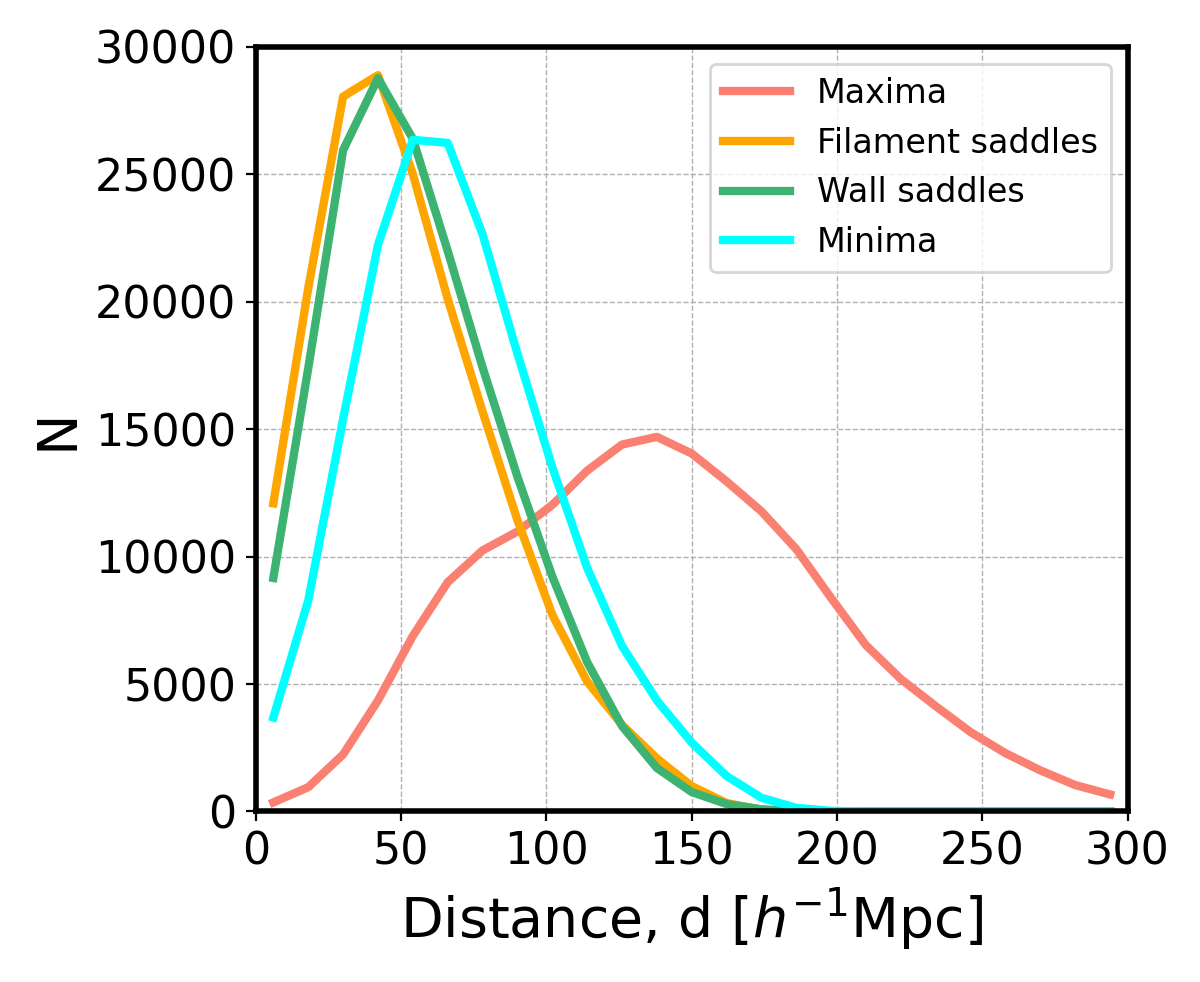}

     \caption{Distribution of minimum distances to the cosmic web critical points obtained from DisPerSE. The histograms show the number of central galaxies as a function of their minimum three-dimensional distance to Maxima, Minima, Filament saddles, and Wall saddles.}
   \label{Diatancesplots-hist}%

  \end{figure}
    
 \begin{figure*}
    \centering
    \includegraphics[width=0.9\columnwidth]{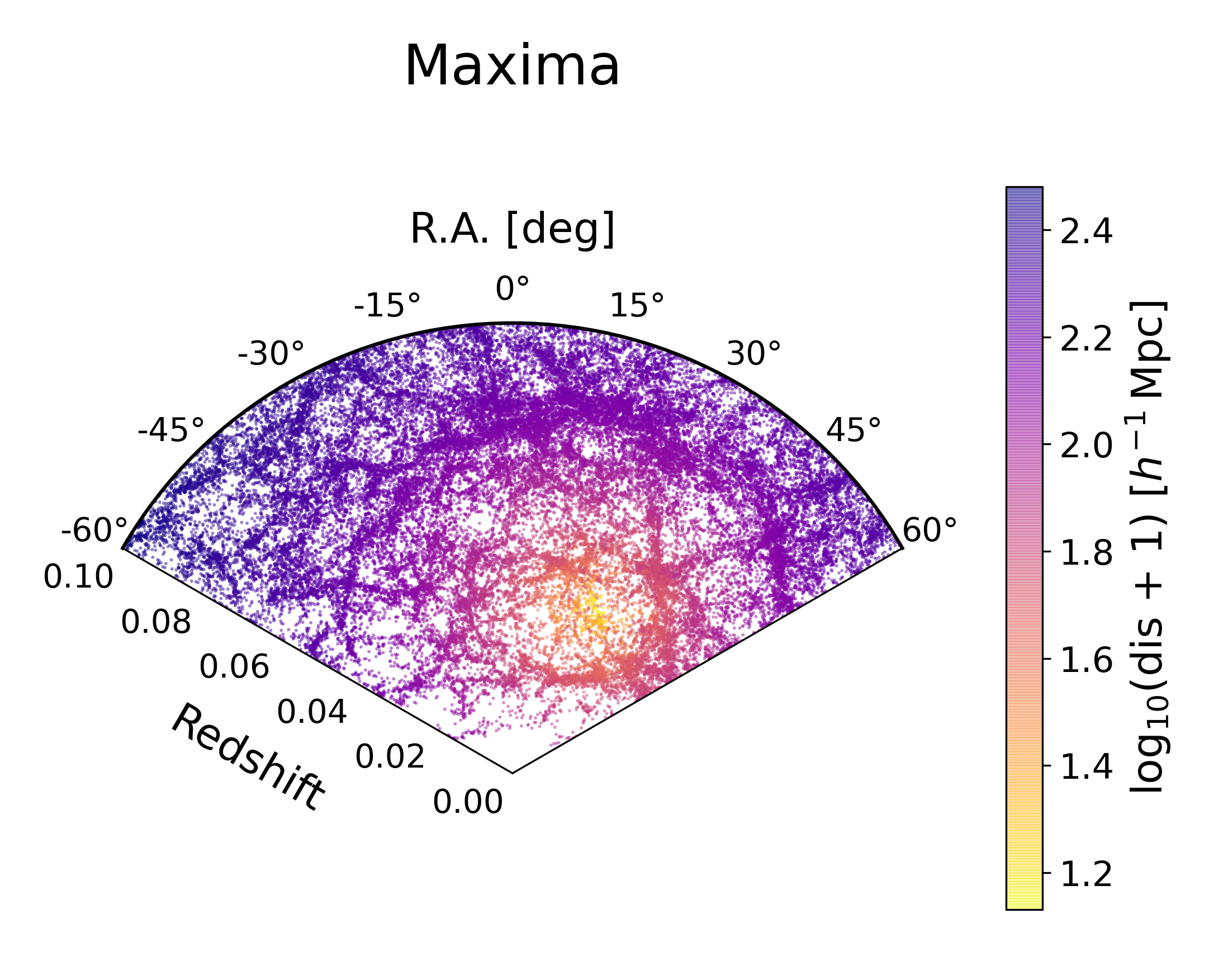}
    \includegraphics[width=0.9\columnwidth]{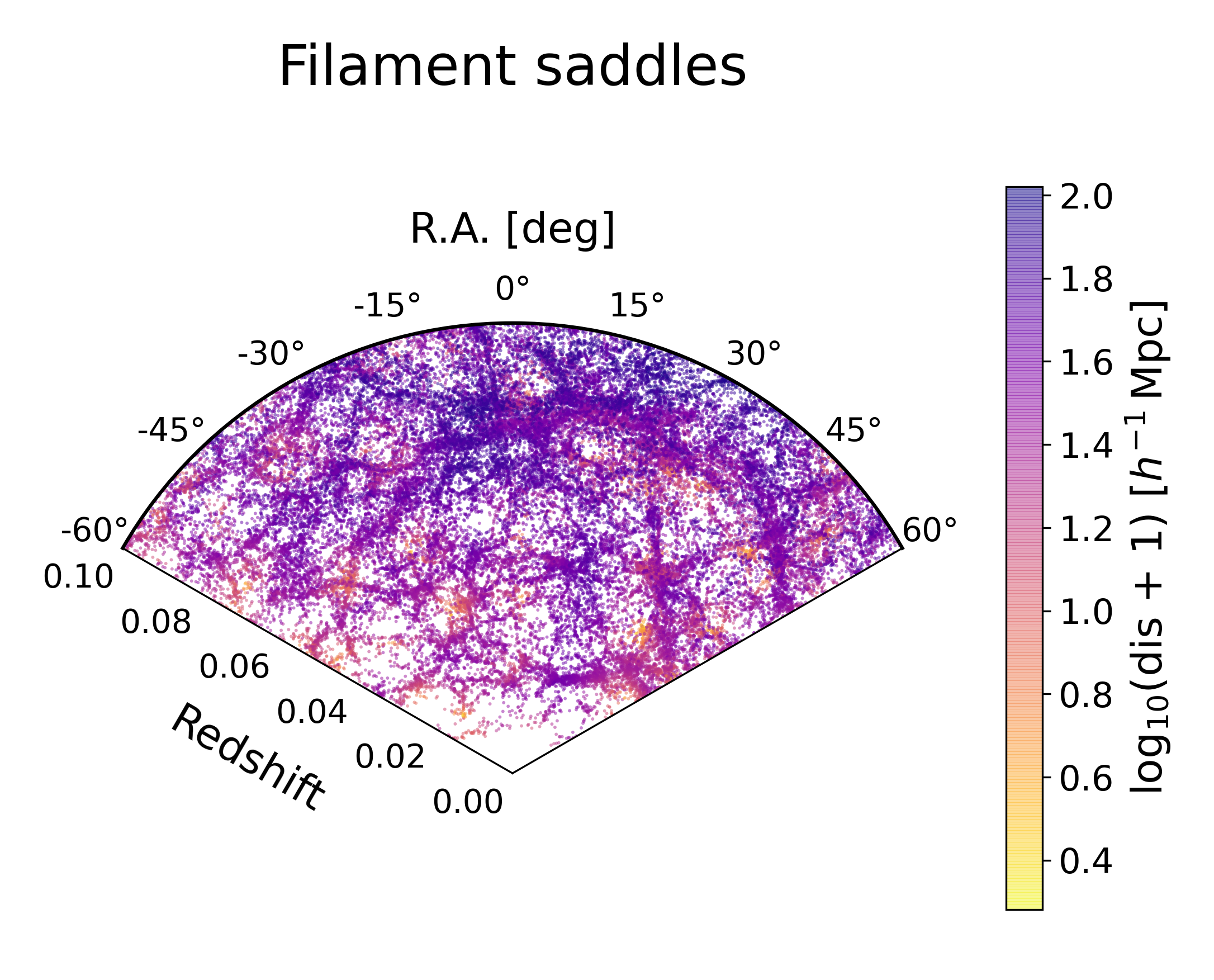}
    \includegraphics[width=0.9\columnwidth]{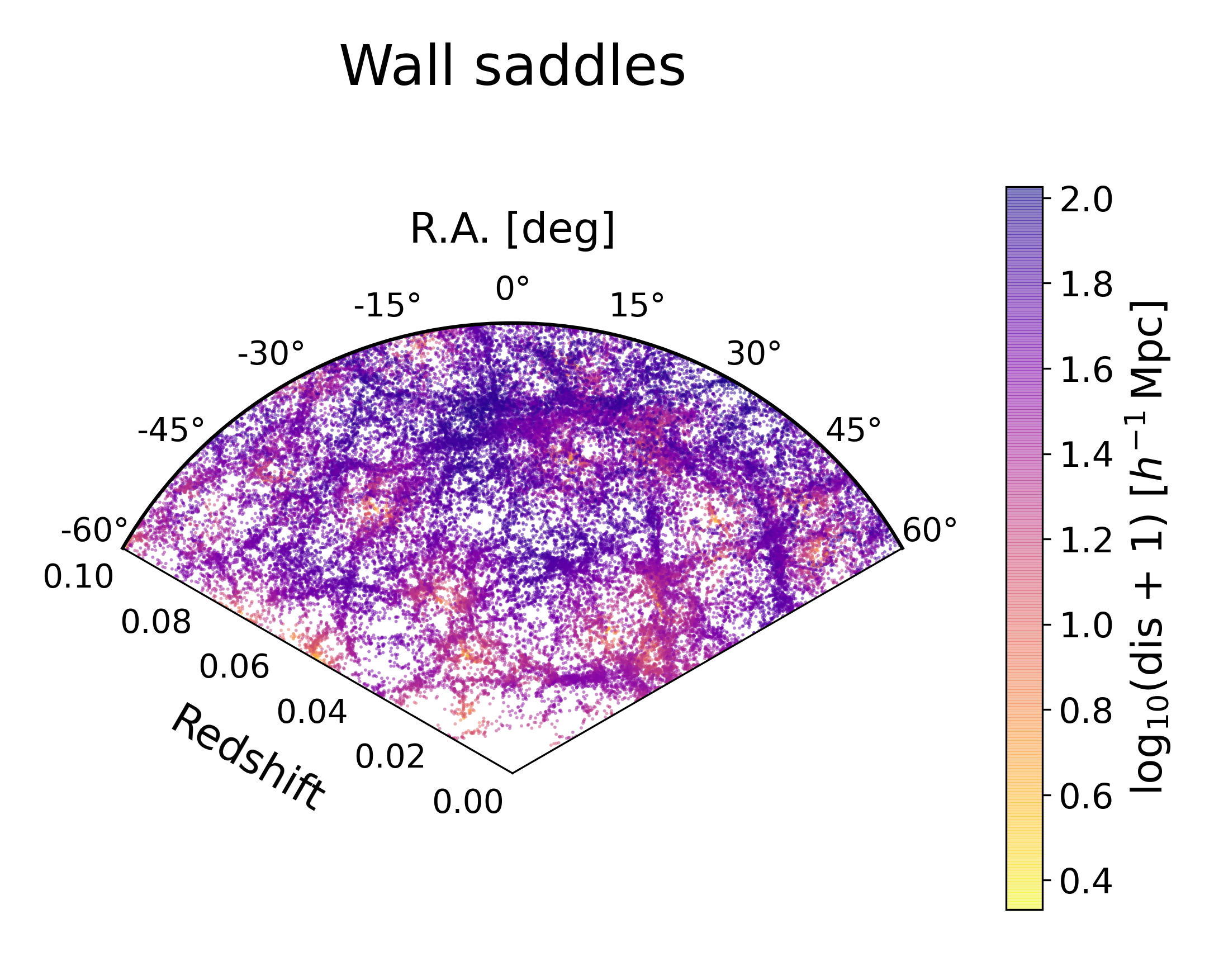}
    \includegraphics[width=0.9\columnwidth]{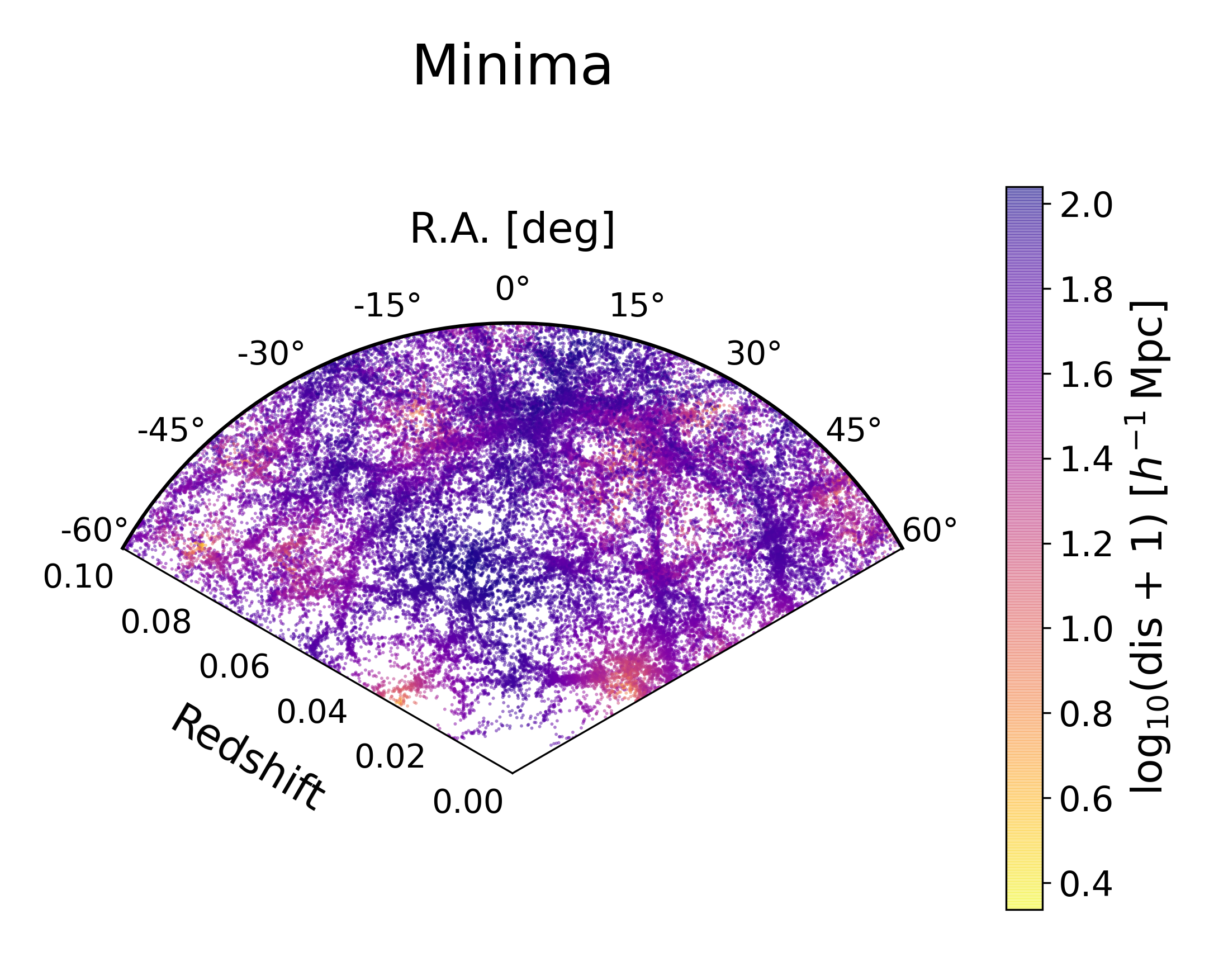}

    \caption{Spatial distribution of central galaxies relative to cosmic web topology. The figure visualizes central galaxies within a $-5^{\circ}$-to-$15^{\circ}$ declination slice of the SDSS Legacy North survey. The panels display the spatial positions of central galaxies, characterized by their minimum three-dimensional distance to the four DisPerSE critical point types (Maxima, Minima, Filament saddles, and Wall saddles). Colour-coding represents the scaled distance to the corresponding critical point, given by $\log_{10}(\text{distance}+1)$.}
    
    \label{Diatancesplots}%
  \end{figure*}

 \begin{figure}
    \centering
    \includegraphics[width=0.9\columnwidth]{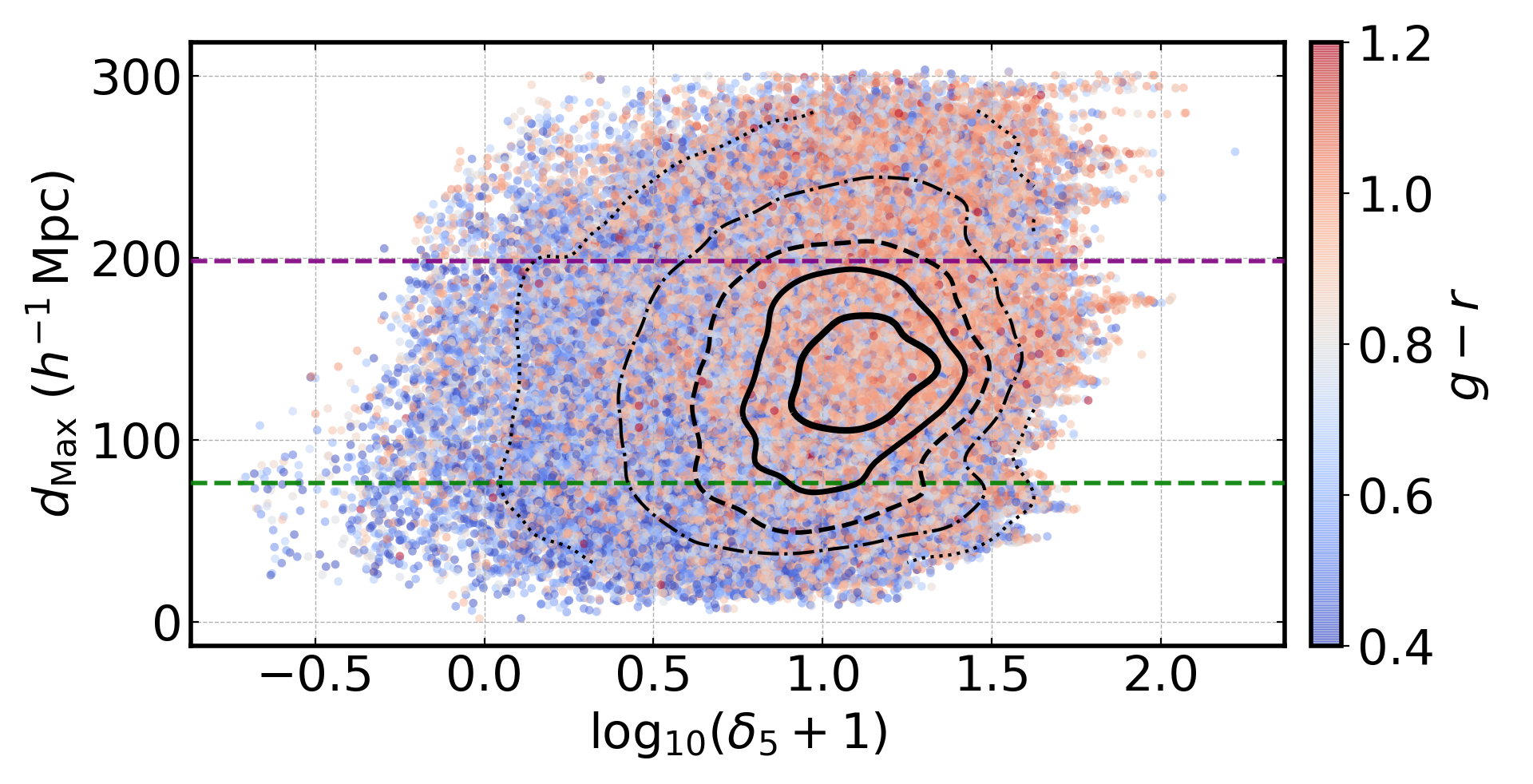}
    \includegraphics[width=0.9\columnwidth]{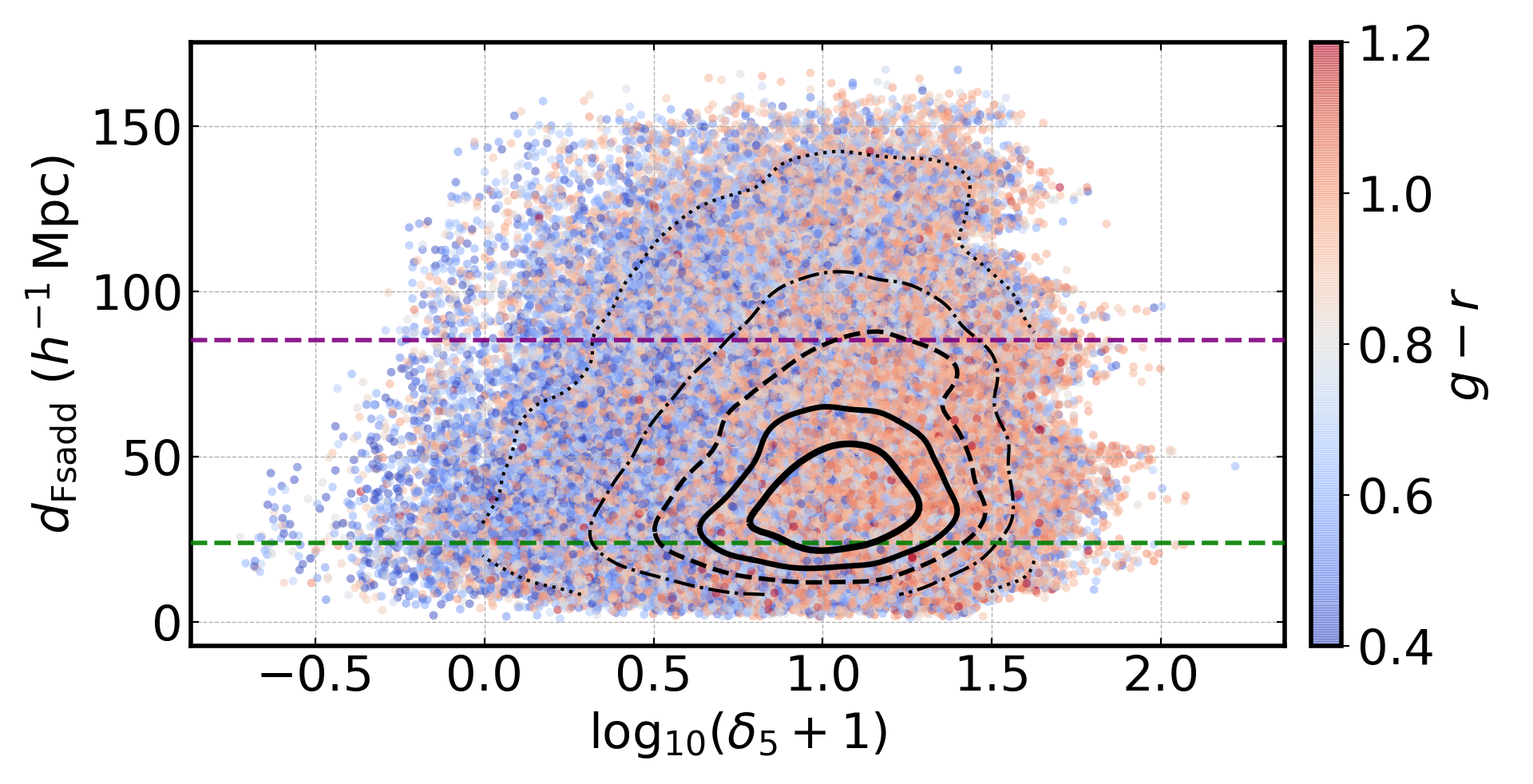}
    \includegraphics[width=0.9\columnwidth]{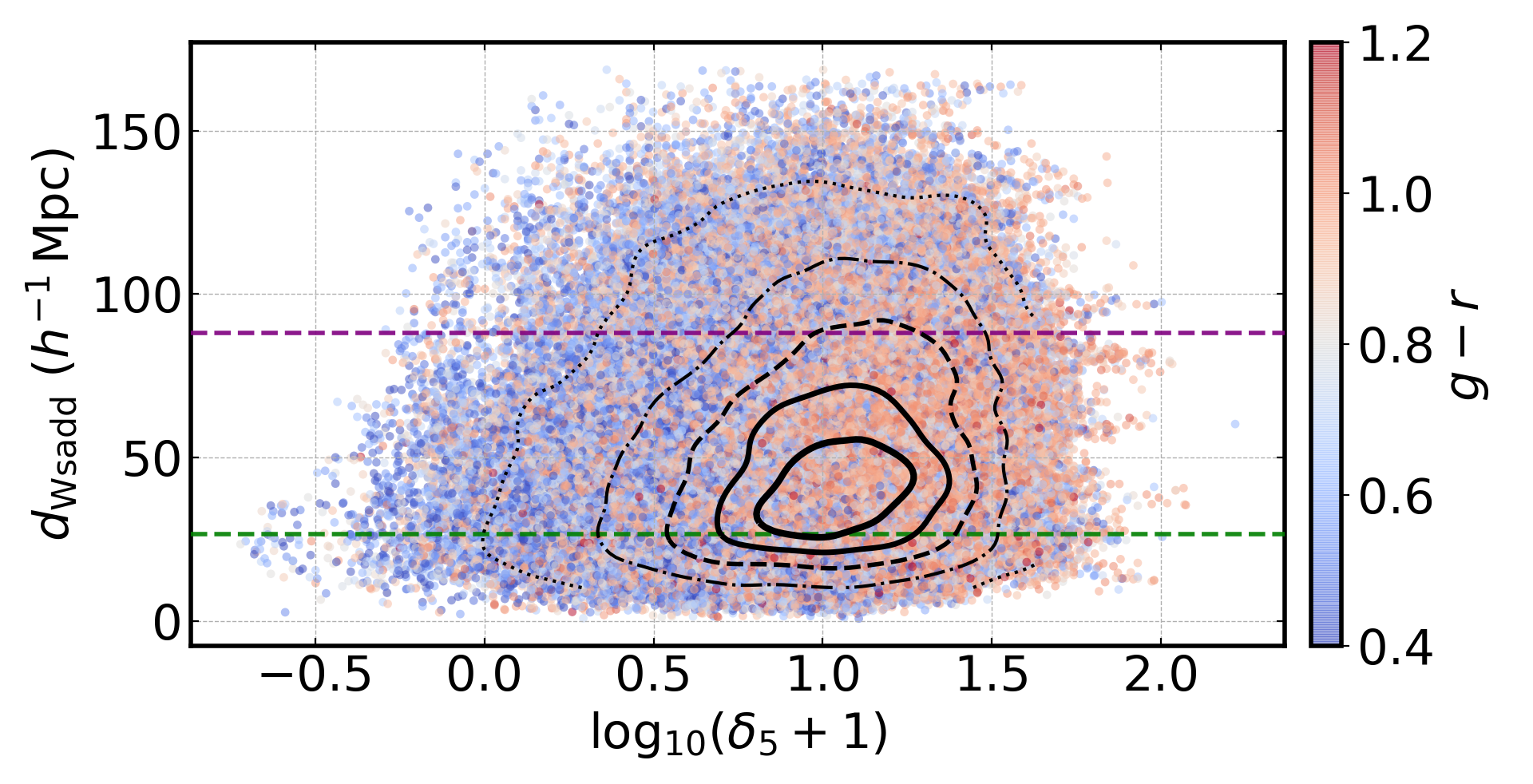}
    \includegraphics[width=0.9\columnwidth]{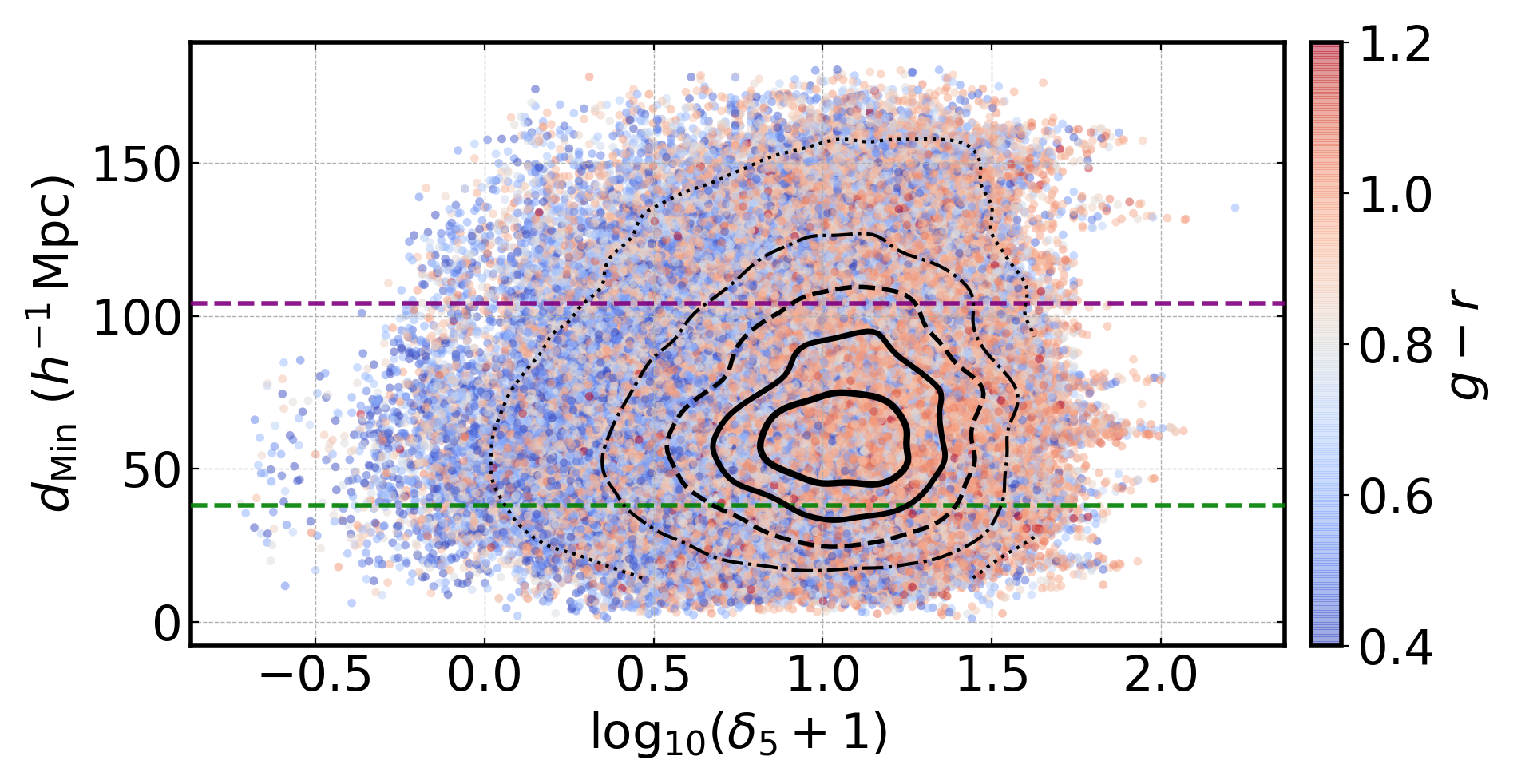}

    \caption{Distribution of central galaxy colour as a function of cosmic web location and local overdensity. In each panel, a scatter diagram shows the relation between the distance to each critical point of the density field and the local group overdensity ($\log_{10}(\delta)$) on $5$ $h^{-1}\, \rm Mpc$ scales for all central galaxies in our sample, colour-coded by their $g-r$ colour. From top to bottom, the panels illustrate the distribution with respect to the distance to: (1) Maxima ($d_{\mathrm{Max}}$), (2) Filament saddles ($d_{\mathrm{Fsadd}}$), (3) Wall saddles ($d_{\mathrm{WSadd}}$), and (4) Minima ($d_{\mathrm{Min}}$). The contour lines (black lines) represent the progressively lower density levels (down to $5\%$ of the peak value). The horizontal dashed lines (green and magenta) mark the 15th ($P_{15}$) and 85th ($P_{85}$) percentiles of the distance distribution, respectively, used to define environment thresholds. }
    
    \label{Dist-density-color}%
  \end{figure}

\section{Data}
\label{sec:data}

\subsection{The SDSS galaxy and group catalogue}
\label{subsec:datasdss}

We base our analysis on the Main Galaxy Sample from the SDSS Data Release 18 \citep[DR18;][]{almeida2023}. This latest data release provides comprehensive multiband photometry (u, g, r, i, z) over a sky area exceeding 10,000 square degrees, complemented by spectroscopic redshifts for over 1.2 million galaxies, extending to z $\sim$ 0.3. The sample's spectroscopic completeness is well-defined for galaxies brighter than an r-band apparent magnitude of 17.77, making it an exceptional resource for statistical studies of cosmic structure. For each object, the catalogue provides precise astrometry, photometry, redshift, and morphological parameters.

The identification of galaxy groups within this dataset was performed using the methodology of \cite{rodriguez2020}, which integrates a Friends-of-Friends (FoF) percolation scheme \citep{merchan2005} with a subsequent iterative halo-based refinement phase \citep{Yang2005}. A key adaptation in our implementation for this work was to extend the selection of potential central galaxies to those with magnitudes as faint as $M_r = -18$. This modification enhances the catalogue's completeness for lower-mass haloes, allowing us to investigate a broader dynamic range of group environments. The algorithm first identifies candidate systems through redshift-space FoF linking. An initial halo mass is then assigned to each candidate based on the total luminosity of its member galaxies. In the iterative phase, these mass estimates are used to refine membership and recalculate halo properties until a stable solution is achieved, robustly identifying systems ranging from poor groups to massive clusters.

The final group catalogue defines galaxy memberships and provides estimates for key halo properties, such as the halo mass ($M_{\text{group}}$), which is derived through abundance matching \citep{Vale2004,Kravtsov2004, conroy2006, behroozi2010}. This technique calibrates the relationship between a group's characteristic luminosity and the mass of its underlying DM halo. Furthermore, galaxies are classified as centrals (the brightest group member) or satellites \citep[e.g.,][]{rodriguez2021}. The reliability of this kind of group identification is well-established: its mass estimates show excellent agreement with weak lensing measurements \citep[e.g.,][]{Gonzalez2021}, and it accurately reproduces galaxy population statistics (e.g., the halo occupancy distribution) from simulations across diverse environments \citep[e.g.,][]{alfaro2022, RodriguezMedrano2023}. For our study, this data set provides the essential spectroscopic, photometric, and group-based properties required to conduct a detailed investigation of secondary halo bias. To ensure a robust cosmic web classification, we restrict our analysis to the central, most complete region of the catalogue, using all groups up to a redshift of  $z = 0.1$, where the large-scale cosmic environments, as determined by DisPerSE, are well-defined. This sample comprises a total of more than 180,000 groups with a mass range of $10^{11.3}$ to $10^{15.0}$ $h^{-1}$M$_{\rm \odot}$.

Throughout this work, we discuss groups, their masses, and related properties, but it is important to note that our analysis of secondary bias primarily relies on the central galaxies (the brightest member of each group). For most groups, especially those of low mass, this refers to a single galaxy, which is by definition the central galaxy.

Figure \ref{Grupdists} displays the mass distribution of the groups analysed in this work (upper panel). This observed distribution follows the characteristic shape of the group mass function, where the number of systems decreases steeply with increasing mass, a well-known manifestation of the hierarchical halo formation paradigm. The bottom panel of Fig. \ref{Grupdists} shows the colour distribution of the central galaxies within these groups, revealing the expected bimodal form that separates the red and blue galaxy subpopulations.

\subsection{Cosmic web characterisation}
\label{subsec:disperse}

This work provides observational constraints on secondary halo bias using central galaxies/groups located in different cosmic environments. The cosmic web is characterised based on the publicly available filament catalogues derived by \citet{Malavasi2020a} using the Discrete Persistent Structures Extractor \citep[DisPerSE;][]{Sousbie2011a,Sousbie2011b}. DisPerSE
identifies persistent topological features (such as filaments, walls, and voids) by applying discrete Morse theory \citep[][and references therein]{Morse1934} to the cosmic density field. The algorithm first computes the density field from the discrete galaxy distribution using the Delaunay Tessellation Field Estimator \citep[DTFE;][]{van2009}. It then locates critical points (Maxima, Minima, and Saddles) where the density gradient vanishes and reconstructs the manifold structure (the Morse complex) that defines the cosmic web. In this framework, the Maxima are typically associated with high-density regions, the saddles correspond to Filaments and Walls, and the Minima are linked to voids or empty cosmic areas. 

The implementation of \cite{Malavasi2020a} adapts DisPerSE to account for the observational limitations and projection effects inherent to the SDSS spectroscopic survey. Their analysis, performed on the SDSS DR12 Main Galaxy Sample, produced multiple catalogues of critical points and filaments by varying the persistence threshold (which controls the significance of detected features) and the smoothing scale applied to the density field \footnote{\url{https://l3s.osups.universite-paris-saclay.fr/cosfil.html}}. These catalogues have been used in several previous studies \citep{Malavasi2020b, Tanimura2020a, Bonjean2020, Tanimura2020b, Izzo2025, Rodriguez2025}. Although we conducted this analysis using DR18, the resulting critical points and structures of the cosmic web are expected to remain consistent with the latest data releases. This consistency is largely due to the minimal large-scale differences observed in the spectroscopic samples. Additionally, the derived critical points should remain largely unaffected by minor variations in data releases. In future analyses, these critical points, which act as stable structural references, can be integrated with galaxy samples from more recent data publications.

In our analysis, we employ the catalogue generated from the Legacy North SDSS galaxy distribution, constructed using a $5\sigma$ persistence threshold and two smoothing cycles. This specific parameter choice is the most conservative and restrictive, ensuring that the identified filaments are highly reliable. A consequence of this strict selection is a significant reduction in the number of detected Minima and, even more so, of Maxima, leaving only a few of these critical points in the final catalogue. The specific combination of persistence threshold and smoothing adopted here was selected based on the extensive analysis of \cite{Izzo2025}, who used these same DisPerSE catalogues to investigate potential variations in the HOD as a function of environment, a probe inherently linked to secondary bias. We have verified that using less restrictive catalogues yields consistent trends, although the amplitude of the relative bias differences decreases due to the inclusion of less prominent and noisy structures.

With this description of the cosmic web, the environment of each central galaxy in the \cite{rodriguez2020} group catalogue was characterised by quantifying its relative location within the geometry of the cosmic web, rather than using a purely categorised environmental label. This was achieved by calculating the minimum three-dimensional distance of each galaxy to the various critical points defined by DisPerSE. These distances are not a direct measure of local density, but rather an indicator of the proximity to the nearest large-scale morphological structure (nodes, filaments, or walls). We specifically focus on four types of critical points: Maxima, Filament saddles, Wall saddles, and Minima, as they represent the most commonly defined cosmic web environments.
Throughout this work, we denote the corresponding distances as 
$d_{\rm Max}$, $d_{\rm Fsadd}$, $d_{\rm Wsadd}$, and $d_{\rm Min}$, 
respectively. 
This distance-based quantitative approach provides a continuous measure of environmental influence. It allows us to classify the structures in a manner directly analogous to the methodology established in the simulation-based work of \cite{MonteroRodriguez2024}.

The distribution of the minimum distance to the DisPerSE critical points
is shown as histograms in Fig. \ref{Diatancesplots-hist}. In addition,  Figure \ref{Diatancesplots} presents the $\log_{10}$-scaled, colour-coded maps tracing how these distances vary across the large-scale structure within a $-5^{\circ}$-to-$15^{\circ}$ declination slice of the SDSS Legacy North survey. The vast distances to Maxima seen in these figures are a direct reflection of our conservative DisPerSE parameter choice, which naturally yields a very low number of nodes.

\subsection{Local group overdensity characterisation}
\label{subsec:overdensity}

We complement our analysis of the environment by measuring the local group overdensity ($\delta$) within spheres of $5$ and $10 \, h^{-1} \, \rm Mpc$ around each central galaxy. This overdensity is calculated by comparing the number of observed galaxies ($N_{\text{gal}}$) to the number expected in a homogeneous universe ($N_{\text{rand}}$). The reference sample $N_{\text{rand}}$ is generated randomly to match the observational mask and the redshift distribution of the SDSS survey. The random catalogue used for this purpose contains more than 50 times the number of galaxies in our observed sample ($N_{\text{total, rand}} > 50 \times N_{\text{total, gal}}$), ensuring minimal Poisson noise in our density estimates. The overdensity $\delta$ is defined as the relative fluctuation about the mean density.
However, as a global normalisation is required, the definition becomes:
\begin{equation}
\delta = \frac{N_{\text{gal}}}{N_{\text{rand}}} \times \frac{N_{\text{total, rand}}}{N_{\text{total, gal}}} - 1
\end{equation}
\noindent where $N_{\text{total, gal}}$ and $N_{\text{total, rand}}$ are the total numbers of galaxies in the observed and random catalogues, respectively.
The measurements corresponding to $5$ and $10 \, h^{-1} \, \text{Mpc}$ radii are hereafter referred to as $\delta_5$ and $\delta_{10}$.
To enhance the statistical robustness of this measurement, all galaxies (both centrals and satellites) are included in the $N_{\text{gal}}$ count. We have verified that our results are qualitatively consistent when only central galaxies are used, demonstrating that our conclusions are not sensitive to this choice.

Figure \ref{Dist-density-color} shows the distance to critical points as a function of local group overdensity $\delta_5$, with each central galaxy colour-coded based on its $g-r$ colour. Upon visual examination, it is evident that there is a very weak correlation between the distances to the various critical points ($d_{\mathrm{Max}}$, $d_{\mathrm{Fsadd}}$, $d_{\mathrm{Wsadd}}$, and $d_{\mathrm{Min}}$) and the local density.  To complement this qualitative assessment, we computed the Pearson correlation coefficient and found that, in all cases, its absolute value is below 0.025, confirming the absence of any significant correlation. However, we acknowledge that even very weak environmental correlations could be non-negligible in the context of halo clustering, as demonstrated by studies such as \cite{Zehavi2018}, \cite{Artale2018}, \cite{MonteroDorta2020B}. Specifically, additional tests show that while no significant trends are observed for the other critical points, the mean values of $d_{\text{Max}}$ can vary by up to $\sim$20\% across the overdensity range considered. Although this trend is subtle and represents the most notable signal among the distances analyzed, such a level of correlation has been shown to potentially induce non-negligible effects on occupancy variations and, consequently, on the resulting clustering signals. In contrast, there is a significant correlation between this local density $\delta _5$ and the $g-r$ colour, as higher densities are systematically related to red central galaxies. As for the relationship between distances to critical points and colour, no clear trend is evident between galaxy colour and proximity to critical points such as minima, filaments, or walls. We have verified that using the local density measured on the scale of $10 \, h^{-1}\,  \rm Mpc$ yields qualitatively similar results. It is important to emphasise again that we are analysing the distances to critical points, not necessarily the environments themselves. This explains why the relationship between these distances and properties such as the colour of central galaxies does not appear to be straightforward. Furthermore, because galaxies of all masses are mixed in Fig.~\ref{Dist-density-color}, both red and blue systems appear across the full density range, including high-density regions. This scatter is amplified by the small number of identified critical points—especially maxima and minima—which provides only a sparse sampling of the cosmic-web skeleton. As a result, galaxies of all colours and densities are found at a wide range of distances from critical points, diluting any clear trends.

\section{Secondary bias measurement}
\label{sec:secondarybias}

To quantify secondary bias in our galaxy group sample, we measure the relative clustering at fixed group mass of several central galaxy (group) subsets defined by both environmental and intrinsic properties. The large-scale environment of each central galaxy is characterised using the distance to various critical points of the cosmic web (Maxima, Minima, Filament saddles, and Wall saddles), as identified by DisPerSE, whereas the density within 5 and 10 $h^{-1} \, \rm Mpc$ is employed as a local environment descriptor. The $g-r$ colour is chosen as the intrinsic central galaxy/group property. 

Central galaxies within each group mass bin are divided into subsets based on percentile rankings according to the aforementioned properties. We explore multiple sample definitions by varying both the percentile thresholds used to define subsets and the specific property ranges examined. This multifaceted approach allows us to probe different manifestations of secondary bias while rigorously controlling for halo mass.

We compute the relative bias $b_{\rm relative}$ between these selected subsets following the approach of \citet{2018Salcedo, SatoPolito2019, MonteroDorta2021, MonteroRodriguez2024}, which relies on cross-correlations with the full sample of groups to maximise signal-to-noise. The relative bias is calculated from the projected cross-correlation functions as:
\begin{equation}
b_{\rm relative}(r_p, S \mid M_i) = \frac{w_{p,[S,\mathrm{all}]}(r_p)}{w_{p,[M_i,\mathrm{all}]}(r_p)},
\end{equation}
where $w_{p,[S,\mathrm{all}]}(r_p)$ is the projected cross-correlation function between central galaxies in the subset selected by property $S$ in a given mass bin $M_i$ and the entire galaxy group sample, and $w_{p,[M_i,\mathrm{all}]}(r_p)$ is the projected cross-correlation between all central galaxies in the same mass bin $M_i$ and the full group galaxy sample. These measurements are performed on scales ranging from  $0.2$ to $30\, h^{-1}\, \rm Mpc$, covering both the one-halo and two-halo regimes where the correlation functions scale approximately as the square of the bias. The final relative bias parameter used to probe secondary bias is, however, obtained by averaging over scales of $5$ to $20$ $h^{-1}\,\rm Mpc$, which is a criterion commonly used to isolate the large-scale effect in the context of secondary bias (e.g., \citealt{SatoPolito2019, MonteroDorta2021, MonteroRodriguez2024}):

\begin{equation}
b_{\rm relative}(S \mid M_i) = \langle b_{\rm relative}(r_p, S \mid M_i) \rangle _{5-20 \,\, h^{-1} \, \rm Mpc}, 
\end{equation}

\noindent The projected cross-correlation function $w_p(r_p)$ is calculated from the observable redshift-space coordinates using:
\begin{equation}\label{eq:wp}
w_p(r_p) = 2 \int_{\pi_{\min}}^{\pi_{\max}} \Xi(r_p, \pi) \mathrm{d}\pi,
\end{equation}
where $\pi_{\min}=0.1 \, h^{-1}\, \rm Mpc$, $\pi_{\max}=20 \, h^{-1} \, \rm Mpc$, and the $\Xi(r_p, \pi)$ estimator is given by:
\begin{equation}
\Xi(r_p, \pi) = \frac{CG(r_p, \pi)}{CR(r_p, \pi)} \cdot \frac{N_{R}}{N_{G}} - 1,
\end{equation}
\noindent and $CG(r_p, \pi)$ and $CR(r_p, \pi)$ represent the central-galaxy and central-random pair counts separated by $r_p$ and $\pi$ (the line-of-sight distances), respectively; $N_{G}$ and $N_{R}$ are the total numbers of galaxies and random points in the sample. The random catalogue was generated by uniformly distributing points across the SDSS footprint, reproducing both the angular and redshift distribution of the SDSS galaxy sample. These computations were performed using the correlation code developed by \citet{Rodriguez2022}, which calculated similar correlation functions for galaxy samples selected by colour, mass, and other physical properties.

It is important to note that the relative bias can itself be interpreted as a measure of the large-scale environment of galaxies, and is therefore expected to exhibit some degree of correlation with the environmental properties discussed here. Throughout this work, we implicitly distinguish between a global clustering property, quantified through two-point statistics, and environmental properties, which are either local, one-point statistics such as overdensities, or metrics associated with specific features of the cosmic web.

\section{Observational constraints on secondary bias}
\label{sec:results}

This section presents a systematic investigation of secondary bias based on the different galaxy and environmental properties described above. We begin by establishing a reference signal using galaxy colour, a well-studied intrinsic property known to correlate with halo assembly history (e.g., \citealt{MonteroDorta2021}). For each property, we measure the clustering of subsamples of central galaxies, divided by their specific characteristics, using narrow group mass bins to ensure a fixed-mass baseline and enable fair comparisons. The mass bins used are $0.2$ dex wide in $\log_{10}(M_{\rm group}/h^{-1}M_{\rm \odot})$ covering the range between $11.3$ and $13.1$, plus two additional higher mass bins spanning $13.1$ to $13.5$ and $13.5$ to $14.3$. 

As mentioned above, we employ a broad definition of secondary bias that includes any potential secondary dependence. Although the environmental distances and local density are not intrinsic halo properties, they may serve as probes of environmentally dependent accretion and formation history. As discussed in the literature based on simulations, a link between assembly bias and these environment-based secondary bias trends can be established, arising from the physical processes that affect halo accretion depending on the specific location of haloes within the cosmic web and the geometry of the surrounding tidal field \citep{Dalal2008, Hahn2009, Paranjape2018,Borzyszkowski2017, Musso2018, MonteroRodriguez2024}.

\subsection{Dependence on galaxy colour}
\label{subsec:color_bias}

We begin our analysis by examining the secondary bias produced by galaxy colour ($g-r$), an intrinsic property known to correlate with star formation history and halo assembly. Previous observational works have mostly focused on the more directly measurable dependence of clustering on colour at fixed stellar mass (e.g., \citealt{LawSmith2017}). However, \cite{Wang2008} reported a significant dependence of group clustering on central galaxy colour at fixed group mass, with groups hosting red centrals clustering more strongly, particularly for less massive haloes. Results at fixed halo mass are well established in hydrodynamical simulations and are consistent with these observational trends (e.g., \citealt{MonteroDorta2020B}). 

\begin{figure}
\centering
\includegraphics[width=1\columnwidth]{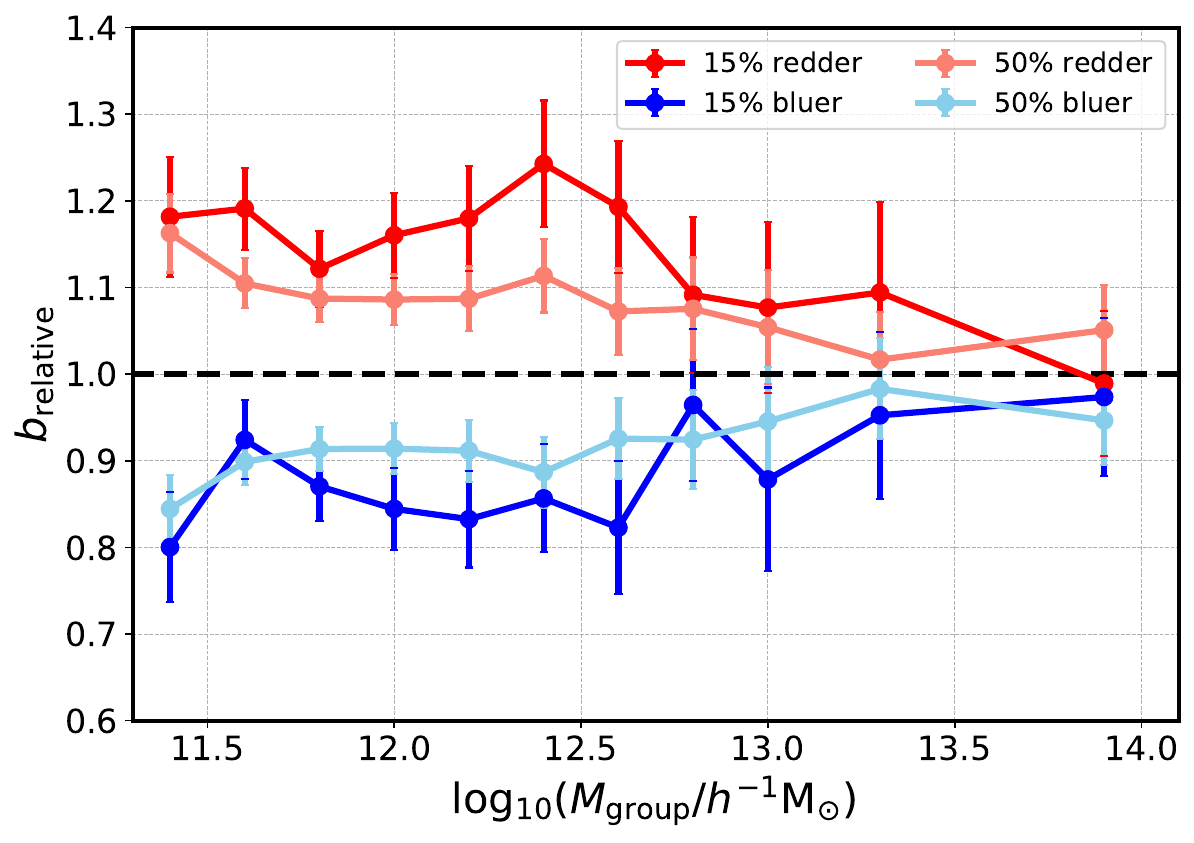}
\caption{Secondary bias measurements for subsets of groups split by the colour of their central galaxies (colour bias). Results are shown for both the extreme 15\% reddest and bluest centrals (darker colours) and the 50\% reddest and bluest centrals (lighter colours).  Uncertainties are estimated using jackknife resampling with 50 subsamples for the correlation functions and propagated to the relative bias.}
\label{fig:SB_color}
\end{figure}

Figure~\ref{fig:SB_color} presents the relative bias for groups split by the colour of their central galaxy, using two selections at fixed halo mass: the 15\% reddest and bluest centrals (saturated colours) and the 50\% reddest and bluest centrals (lighter colours). Throughout this figure and the remainder of our analysis, uncertainties are estimated using the jackknife resampling technique, dividing the sample into 50 subsamples for the calculation of correlation functions and error propagation for $b_{relative}$ calculations. We find that groups with redder centrals exhibit systematically stronger clustering than those with bluer centrals. The amplitude of this effect is largest at low halo masses, reaching $b_{\rm relative} \sim 1.2$ for the 15\% reddest centrals and $b_{\rm relative} \sim 0.8$ for the bluest ones, which is consistent with expectations. While the signal is strongest for the extreme 15\% colour selection, it is also noisier due to the smaller sample size. The 50\% split reduces the amplitude slightly, as expected, but provides a significantly cleaner measurement. This secondary bias is most pronounced below $\log_{10}(M_{\rm group}/h^{-1}M_{\rm \odot}) \sim 13.5$ and diminishes at higher masses, a trend consistent with both observational results and simulations \citep{Wang2008, MonteroDorta2020B}. Furthermore, the amplitude and mass dependence of the signal agree remarkably well with the simulation results from Fig. 5 of \citet{MonteroDorta2020B}, showing very similar bias values across the relevant mass range. Despite inherent mass-assignment uncertainties, which could systematically shift or add noise to the signal, the clear trend we recover remains physically meaningful and consistent with simulations. 

Our analysis confirms and extends the early findings of \cite{Wang2008} to a broader mass range, while showing consistent results in the overlapping mass regime. This demonstrates that our sample is capable of directly recovering the colour bias reported in both observations and simulations. These results highlight the role of central galaxy colour as a sensitive tracer of secondary bias, encapsulating the interplay between star formation history, halo assembly, and the large-scale environment. The detection of this effect in our sample provides a benchmark against which other secondary dependencies can be tested.

\subsection{Dependence on cosmic web location and environment}
\label{subsec:cosmic_web_bias}

\begin{figure*}
    \centering
	\includegraphics[width=2\columnwidth]{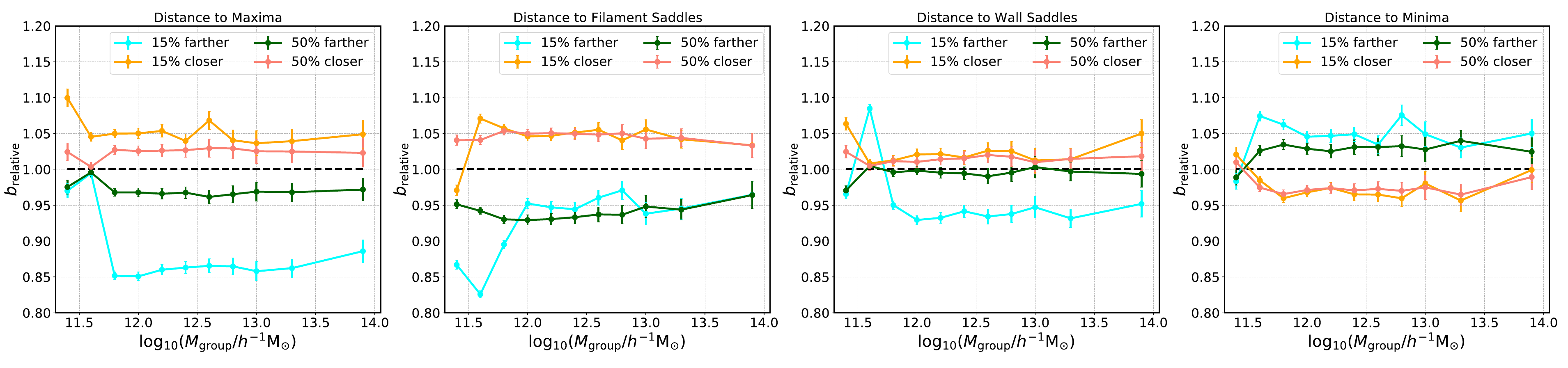}
\caption{Secondary halo bias based on the distance to the critical points of the cosmic web. Each panel shows the relative bias ($b_{\text{relative}}$) as a function of the logarithm of the group mass ($\log_{10}(M_{\rm group}/h^{-1}M_{\rm \odot})$). The panels are organized into two rows and four columns, with each column representing a specific cosmic web feature identified by the DisPerSE algorithm: Maxima (first column), Filament saddles (second), Wall saddles (third), and Minima (fourth). In each panel, the cyan and orange lines represent the 15\% of galaxies farthest from and closest to the feature, respectively. The green and red lines represent the 50\% of galaxies farthest from and closest to the feature, respectively.  In each panel, the orange line represents the subset of galaxies closer to the feature, and the light blue line shows the subset of galaxies farther away. The dashed black line at $b_{\rm relative}=1.00$ serves as a reference point for the relative bias. Error bars on each data point were calculated using the Jackknife method.}

    \label{fig:SB_distances}
\end{figure*}

\begin{figure*}
    \centering
	\includegraphics[width=1.5\columnwidth]{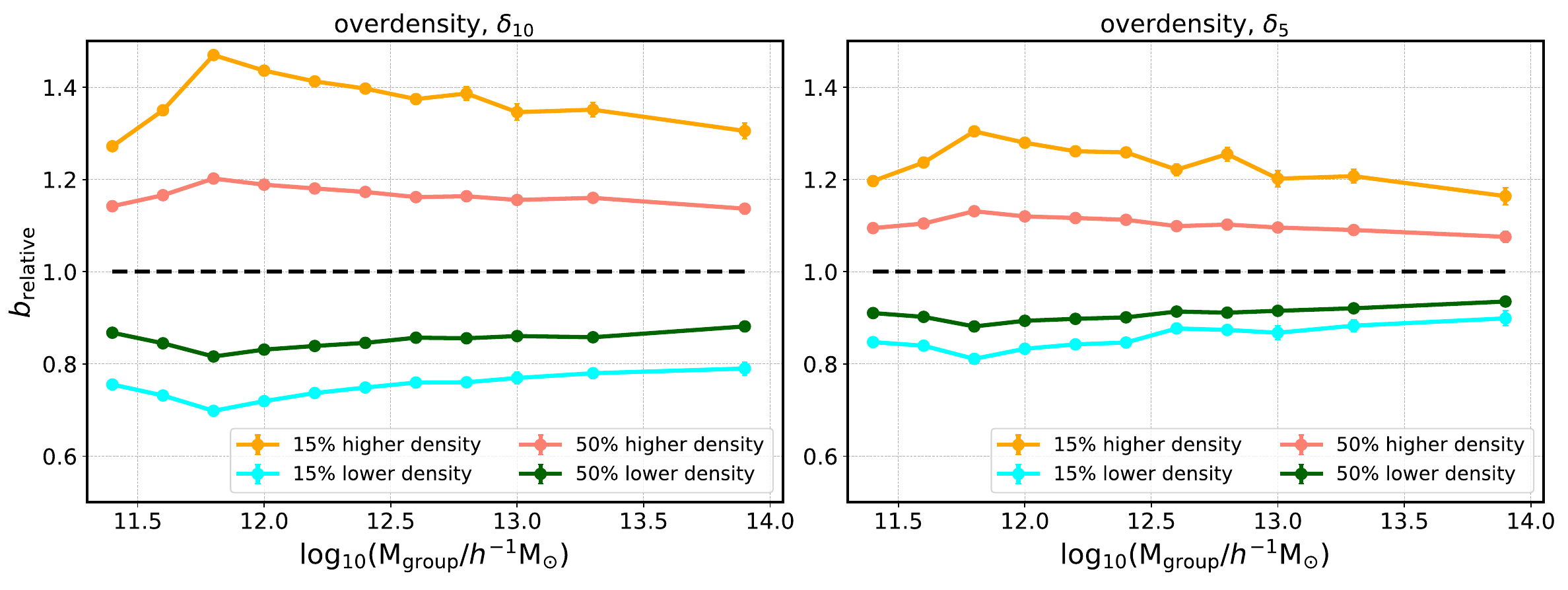}
    \caption{Secondary bias based on local group overdensity presented in the same format of Fig. \ref{fig:SB_distances}. The four lines represent galaxy sub-samples split by their local density, with two lines for higher densities (15\% and 50\%) and two for lower densities (15\% and 50\%). The error bars on each data point were estimated using a jackknife resampling method.}
    \label{fig:SB_density}
\end{figure*}

Building on the colour analysis, we now turn to the analysis of an environmental factor: the specific location within the cosmic web, which has been shown to influence group and galaxy properties (e.g., \citealt{Einasto2024}). We investigate how secondary bias relates to this cosmic web environment by measuring the clustering dependence on the distance to its critical points.The results of our analysis are presented in Fig. \ref{fig:SB_distances}. The figure shows the relative bias as a function of the group mass for populations split by their distance from the four critical features identified by DisPerSE: Maxima, Filament saddles, Wall saddles, and Minima. Here, the distance from the closest point among these categories is adopted. 

A primary result, consistent with \cite{MonteroRodriguez2024}, is that the distance parameters induce significant secondary bias, although the nature of the signal is highly dependent on the specific feature. The analysis reveals a clear difference when comparing 
the groups with the 15$\%$ highest and lowest distance to the Maxima, where the farthest population exhibits a relative bias $b_{\text{relative}} < 0.89$ in all the range bins. When using the median split (50\% closest and farthest), the differences remain pronounced, but exhibit greater symmetry; the populations farthest from the features show $b_{\text{relative}} \sim 0.97$, while the closest show $b_{\text{relative}} \sim 1.03$. This symmetric trend is evident across the entire mass range. 

For Filament saddles, the observed trend aligns with that of the Maxima, but exhibits a key difference: the bias signals for the 15\% and 50\% subsets are more consistent with each other. The relative bias stabilises at approximately 1.05 for haloes closest to the filaments and 0.95 for those farthest away.

An analysis of Fig. \ref{fig:SB_distances} reveals that the secondary bias signal associated with Wall Saddles is notably weaker than that of Maxima or Filament Saddles. A consistent signal, comparable to that of Filament Saddles, is only observed for 15\% of groups farthest from the Wall Saddles. For all other cases, particularly when considering groups closest to these features or when using the 50\% split, the relative bias remains close to unity, showing no significant effect. This overall subdued signal stands in contrast to the strong secondary bias predicted for Wall Saddles in \cite{MonteroRodriguez2024}.

The results for the Minima align with the expected theoretical framework, yet reveal important quantitative details. We find a symmetric signal where groups closer to the voids exhibit a relative bias of approximately 0.97, while those farther away show a bias of about 1.03. This pattern is remarkably consistent for both the 15\% and 50\% splits, underscoring the robustness of the measurement. However, this symmetric behaviour presents a nuance when compared to \cite{MonteroRodriguez2024}; in that finding, the signal was less balanced, as groups far from the minima displayed a relative bias much closer to unity, lacking a significant signal of enhanced clustering.

These results support a physical interpretation where the clustering strength of haloes, at fixed mass, is modulated by the anisotropy of their large-scale environment as traced by the cosmic web skeleton. Haloes near the most anisotropic features (Maxima and Filament saddles) show enhanced clustering, while those near Minima exhibit reduced clustering. The complex behaviour of Wall Saddles, where only the 15\% most distant groups show a signal comparable to filaments, suggests their environmental influence is more subtle and may represent a transitional regime between isotropic voids and anisotropic filaments. This nuanced picture contrasts with the stronger secondary bias predicted for all these environments in \cite{MonteroRodriguez2024}. Further investigation is needed to confirm that these results are not influenced by systematics in the identification of the cosmic web components.  

Our observational results reveal an important distinction in secondary bias signals: the amplitude of the environmental dependence based on distance to cosmic web critical points, while clearly detectable, is generally weaker than the colour bias reported in Sect. \ref{subsec:color_bias}. This hierarchy is particularly noteworthy given that this signal exhibits a remarkably constant amplitude across the entire mass range studied. This behaviour presents a double contrast: first, with the colour-based secondary bias, whose strength diminishes at higher masses; and second, with predictions from hydrodynamical simulations like those from \cite{MonteroRodriguez2024}, which found a pronounced declining mass dependence in analogous environmental measures. Our results suggest a distinct hierarchy in observational data: internal properties like colour dominate secondary bias, while the influence of the cosmic web environment, though robustly detected, is weaker and mass-independent. This may indicate that this signal is diluted or modulated in observational datasets, potentially due to the challenges in accurately reconstructing the cosmic web from galaxy surveys. 

Finally, we examine the relationship between secondary bias and local group overdensity, a key environmental property that captures the immediate surroundings of each group. This local density measurement complements the large-scale environmental information from cosmic web distances, providing a more complete picture of how environment impacts secondary bias. Figure~\ref{fig:SB_density} shows results for both 15\% and 50\% subsets selected by density. A clear and robust trend emerges across both selections: groups in higher-density environments exhibit significantly stronger clustering, reaching relative bias values up to 1.4 for the 15\% highest density regions ($\delta_{10}$) and 1.3 for $\delta_{5}$. Conversely, groups in lower-density environments show systematically weaker clustering, with $b_{\rm relative}$ values dropping below 0.8 ($\delta_{10}$) and 0.7 ($\delta_{5}$) for the 15\% least dense regions. The signal displays a weak mass dependence and shows a clear amplitude increase when moving from the 50\% to the 15\% selection. Notably, these density-based measurements produce the strongest secondary bias signal in our analysis, surpassing even the well-established colour-based bias. The consistency of these results is particularly meaningful for the $\delta_5$ measurement, where the density is computed within a 5 $h^{-1}\,\rm Mpc$ radius that does not overlap with the scales used for bias calculation (5-20 $h^{-1}\,\rm Mpc$), confirming this as a genuine environmental effect rather than a scale-dependent artifact. We note that \citet{MonteroRodriguez2024} also identify overdensity as the environmental property showing the strongest secondary bias dependence (see their Appendix), although with a smaller overall amplitude. Moreover, while our overdensity- and distance-based signals are weaker than those reported in that work, the colour-based secondary bias in our measurements reaches amplitudes comparable to those found in \citet{MonteroRodriguez2024}.

The larger error bars in Fig. \ref{fig:SB_distances} compared to Fig. \ref{fig:SB_density} reflect a physical finding: the environmental classification based on the cosmic web geometry is noisier or less efficient at segregating halos than that based on local density. This implies a greater intrinsic variance in secondary bias when accounting for the geometry of the environment.

Our analysis reveals a definitive hierarchy in secondary bias drivers. The strongest signal originates from local group overdensity, followed by the secondary bias based on galaxy colour or colour bias. The influence of distance to cosmic web critical points, while clearly detectable, constitutes the weakest of the three effects. This establishes that local environmental measures and internal properties are more dominant determinants of secondary bias than the large-scale cosmic web structure in our observational sample. 

\subsection{Signal significance analysis}

A potential concern when measuring secondary bias in finite mass bins is that the intrinsic dependence of clustering on mass can be can be confused with genuine secondary bias signals. Even within a narrow mass bin, differences in the mean mass of two subsamples will translate into an expected difference in their clustering strength due to the primary halo bias relation. The following analysis quantifies whether our measured secondary bias signals are significantly larger than the differences expected from this residual intra-bin mass spread. Note, however, that this procedure neglects the potential impact of halo mass uncertainties, so it must be regarded as a first-order theoretical test.

\begin{figure}
    \centering
	\includegraphics[width=1\columnwidth]{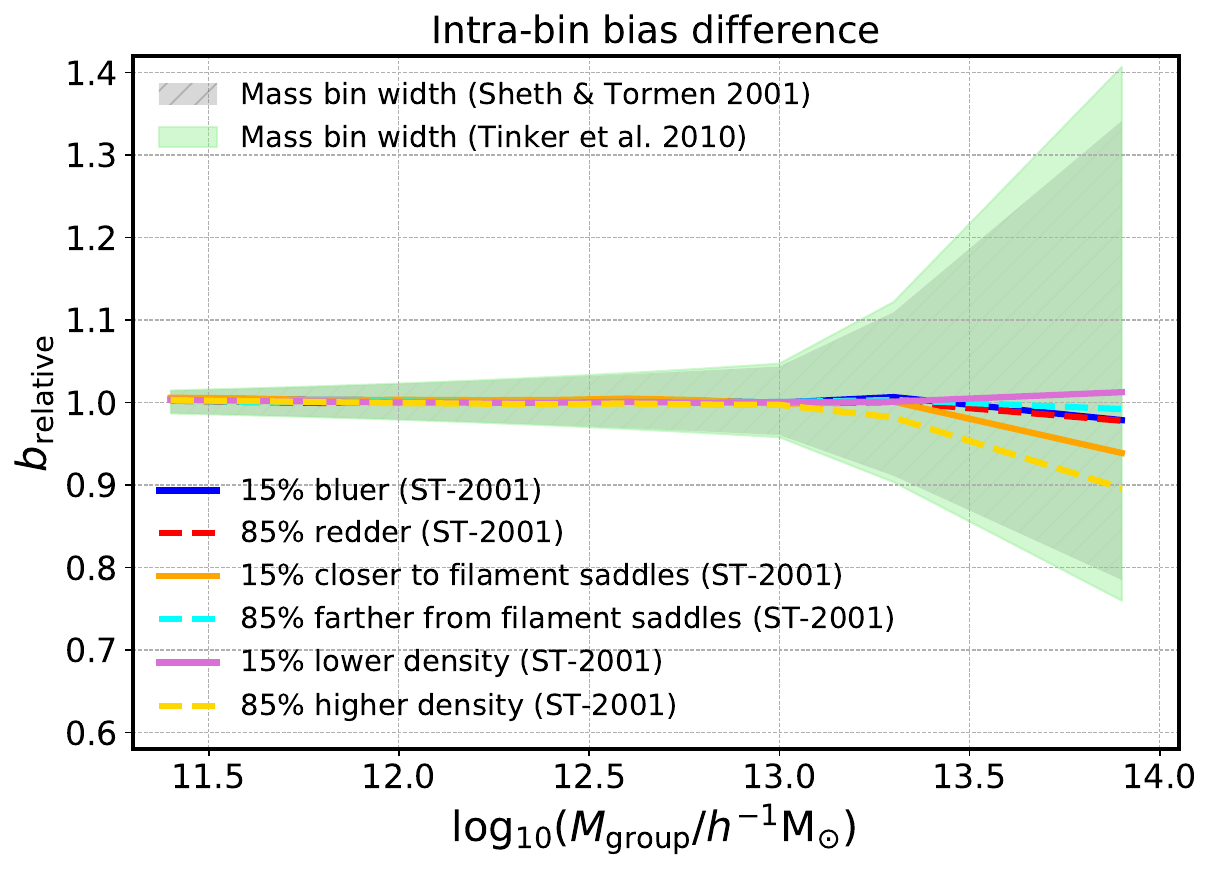}
    \caption{An assessment on the robustness of the secondary bias measurements against potential intra-bin mass dependencies of group bias. Each line corresponds to the relative bias of upper/lower subsets computed assuming the \cite{Sheth2001} halo bias model, where the mass assigned to each subsample corresponds to the mean mass. For reference, the shaded gray and green regions show the maximum intra-bin bias difference, obtained from the edges of the group mass bin; here both the \cite{Sheth2001} model and the  \cite{Tinker2010} analytical fit are employed.}
    \label{fig:SB_diff}
\end{figure}

Figure~\ref{fig:SB_diff} assesses the robustness of our secondary bias signals against potential artifacts from the finite width of our mass bins. We quantify the expected bias variation arising from intra-bin mass spread using established bias models. The colored lines represent the predicted relative bias for each subsample, computed by applying the theoretical \cite{Sheth2001} bias model to each subset's mean mass. For reference, the shaded regions show the maximum theoretical bias difference across the entire mass bin, obtained from both the analytical \cite{Sheth2001} model and the simulation-calibrated \cite{Tinker2010} fit evaluated at the bin edges.

The key result is that the bias differences measured in our analysis (shown in previous sections) are substantially larger than the intra-bin mass-dominated bias differences derived from theoretical models. The signals derived from the subsets' mean masses (colored lines) are significantly weaker than our measurements, being consistent with 1 for most of the mass range. Even the maximum possible variation from the full bin width (shaded regions) would be insufficient to explain our observed signals below \(\log_{10}(M_{\rm group} / h^{-1} M_{\rm \odot}) \lesssim 13\). This confirms that our detected secondary bias cannot be attributed to residual mass dependencies within the bins. We note that the increase in the theoretical expectations at the highest mass bins is due to the larger bin widths employed in that regime.

\section{Discussion and conclusions}
\label{sec:conclusions}

This work provides detailed observational constraints on secondary halo bias: the dependence of halo clustering strength on properties beyond halo mass. We constructed an extended galaxy group catalogue from the SDSS-DR18 specifically for this analysis, applying a refined version of the \citet{rodriguez2020} group-finder algorithm to improve the recovery of low-mass systems and thereby extend the mass range of our determinations. Using this catalogue, our results clearly confirm a key prediction from cosmological simulations and analytical models of structure formation: the large-scale clustering of haloes, traced here by galaxy groups, is significantly influenced by secondary properties at fixed halo mass \citep[e.g.,][]{Gao2007, Dalal2008, Wechsler2018, Paranjape2018, SatoPolito2019, MonteroRodriguez2024}.

We began by evaluating secondary bias based on central galaxy colour, a well-established intrinsic property. Our analysis, conducted across a broader mass range than typically explored in observational studies—more comparable to the dynamic range accessible in simulations—reveals that groups with redder centrals exhibit systematically stronger clustering than their bluer counterparts at fixed mass. This signal is particularly pronounced at lower masses (\(\log_{10}(M_{\rm group} / h^{-1} M_{\rm \odot}) \sim 11.5\text{--}12.7\)). We quantify this effect by computing the final relative bias parameter from scales of 5 to 20 \(h^{-1} \, \rm Mpc\), a conventional range for isolating secondary bias signals. This robust detection confirms and extends the early observational findings of \citet{Wang2008} to a broader mass range, while showing consistent results in the overlapping mass regime. The observed trend aligns with expectations from hydrodynamic simulations \citep{Li2006, MonteroDorta2020B}, which show similar colour-dependent clustering at fixed halo mass. Since central galaxy colour correlates with halo assembly history \citep{Hearin2013,Wechsler2018, Behroozi2019, ChavesMontero2020, MonteroDorta2021}, this observed dependence provides an important observational test of halo assembly bias, suggesting that color bias may serve as an accessible proxy tracing the formation and evolutionary pathways of haloes, even if not directly equivalent to assembly bias itself.

We performed the first detailed observational measurement of secondary bias based on the distance to the cosmic web critical points, identified using the DisPerSE algorithm. Our results reveal a clear environmental dependence: groups of identical masses located near high-density regions (Maxima and Filament saddles) tend to exhibit enhanced clustering (specially when $15\%$ subsets are employed to optimise the signal), while those in underdense regions (voids) or distant from the major structures show suppressed clustering. This demonstrates that the geometric location of groups within the cosmic web fundamentally influences their clustering beyond halo mass, supporting the scenario where haloes residing in different cosmic web environments originate from distinct initial conditions, leading to bias variations even at fixed mass \citep[e.g.,][]{Hahn2009,Borzyszkowski2017,Musso2018, Paranjape2018, MonteroRodriguez2024}.

Our observational results show both significant agreement and notable discrepancies with the predictions from the TNG300 simulation \citep{MonteroRodriguez2024}. As shown in Figure~\ref{fig:SB_distances}, we confirm the general theoretical expectation that cosmic web location modulates secondary bias: haloes near anisotropic features (Maxima and Filament saddles) exhibit enhanced clustering ($b_{\mathrm{relative}} > 1$), while those in isotropic underdense regions (near Minima) show suppressed clustering ($b_{\mathrm{relative}} < 1$). However, two key differences emerge from our observational measurements. First, we find that the secondary bias signal for Wall Saddles is substantially weaker in observations than predicted by simulations. While \cite{MonteroRodriguez2024} identified wall-type saddles as the most powerful environmental discriminators, our measurements show the relative bias for groups segregated by distance to Wall Saddles remains consistent with unity across most of the mass range, with only the 15\% most distant groups showing a modest signal. Second, the mass dependence of environmental secondary bias appears flatter in our observational data compared to the pronounced decrease with mass seen in simulations. These discrepancies could arise from systematic limitations in cosmic web identification within galaxy surveys, particularly for wall-like structures that are challenging to reconstruct from discrete, redshift-space galaxy distributions. To rigorously distinguish between physical origins and methodological artifacts, future work will require mock catalogues that replicate our observational setup, enabling direct comparison with simulation predictions while accounting for survey-specific systematics.

Complementing the cosmic web analysis, we find that local group overdensity produces the strongest secondary bias signals in our study (Figure~\ref{fig:SB_density}). Groups in high-density environments exhibit dramatically enhanced clustering, with relative bias reaching up to 1.4 for the 15\% highest density regions, while those in low-density environments show suppression down to $b_{\mathrm{relative}} \sim 0.7$. This local density signal, measured on scales ($5 h^{-1}\rm Mpc$) that do not overlap with our bias calculation range, represents a genuine environmental effect that surpasses even the well-established colour-based secondary bias. The exceptional strength of local density as a secondary bias indicator aligns with findings from hydrodynamical simulations \citep[e.g.,][]{Borzyszkowski2017,Paranjape2018, Ramakrishnan2019, MonteroRodriguez2024} and theoretical expectations, given that local environment is a fundamental determinant of structure formation and bias. While the local density signal broadly mirrors the environmental trends observed with cosmic web distances—with dense regions showing enhanced clustering and underdense regions suppression—its significantly larger amplitude underscores local density as a more direct and powerful environmental factor in shaping halo clustering, operating somewhat independently of the topological cosmic web framework.

\begin{figure}
    \centering
	\includegraphics[width=1\columnwidth]{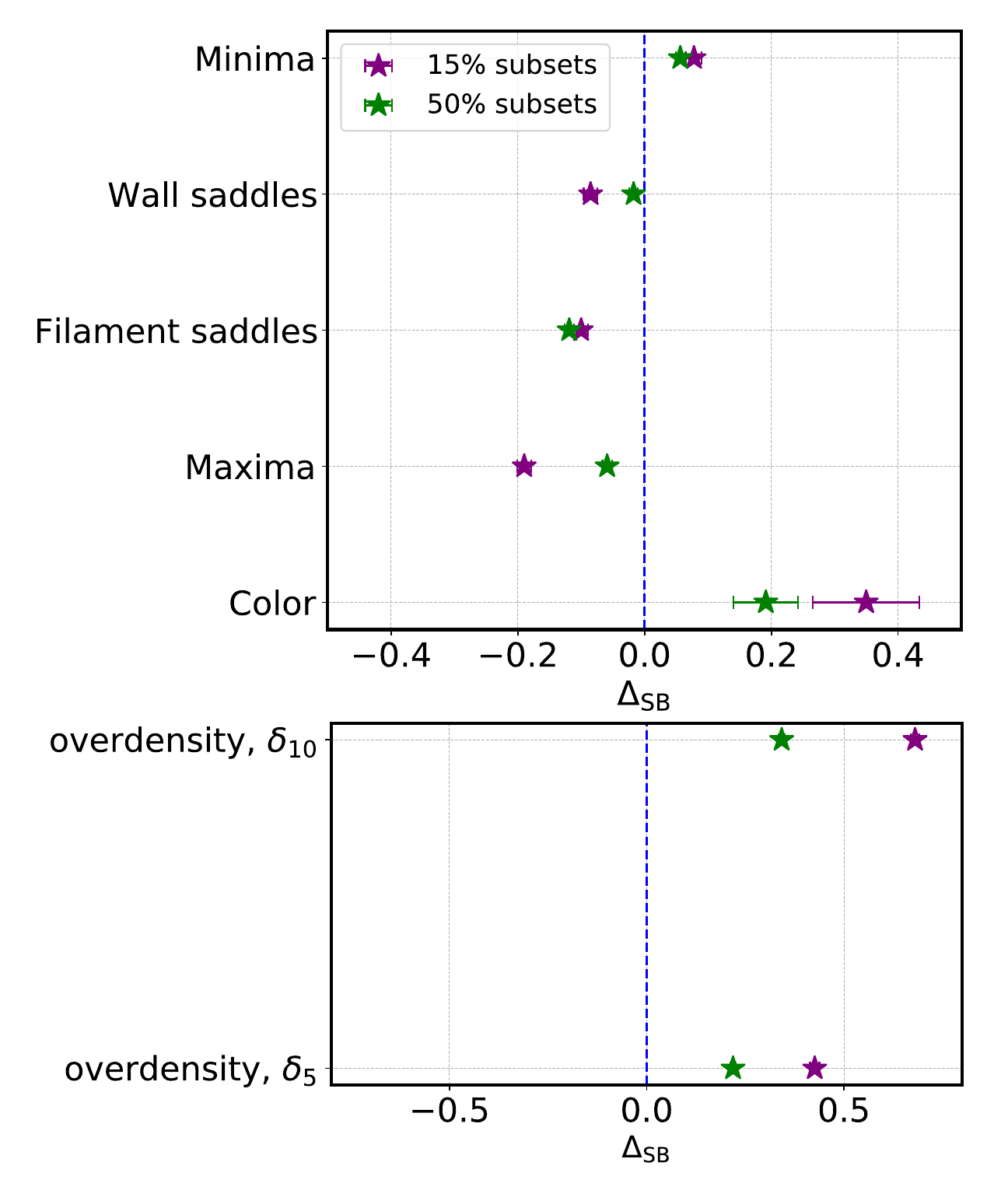}
    \caption{ Comparison of the secondary bias signal for different properties. The stars represent the difference between the average relative bias of the upper subsets and the average relative bias of the lower subsets, in the mass range of $\log_{10}(M_{\rm group}/h^{-1}M_\odot)$ 12 and 12.5. The results are shown for the 15\% (purple) and 50\% (green) subsets. The properties analysed are: local density, Minimum density, Wall saddles, Filament saddles, Maximum density, and colour.}
    \label{fig:summary}
\end{figure}

Our results establishes a clear hierarchy in observational secondary bias. In Figure~\ref{fig:summary}, we compare the strength of several secondary bias signals using the average difference in relative bias between the upper and lower subsets within the representative mass range $\log_{10}(M_{\rm group}/h^{-1}M_\odot) = [12, 12.5]$. This estimator, $\Delta_{\rm SB}$, provides a direct way to quantify and visualise the hierarchy among the different secondary bias effects. The largest integrated signal corresponds to the local overdensity measured on a $10 \,h^{-1}\,\mathrm{Mpc}$ scale ($\delta_{10}$), with $\Delta_{\rm SB} \simeq 0.6$ for the 15\% subset. The overdensity on a smaller scale, $5\,h^{-1}\,\mathrm{Mpc}$ ($\delta_5$), follows with a similarly strong but slightly smaller signal ($\Delta_{\rm SB} \simeq 0.3$). Central galaxy colour also shows a clearly positive signal ($\Delta_{\rm SB} \simeq 0.35$ for the 15\% subset). The geometric quantities (distances to Minima, Wall saddles, Filament saddles, and Maxima) exhibit comparatively weaker signals, with values ranging from mildly positive to mildly negative, and with Wall saddles remaining consistent with zero. The overall hierarchy is robust, as the ordering of signal strengths is consistent between the 15\% and 50\% subsets. 
We emphasise that, as with any observational study, we have the inherent uncertainty of mass assignment that could affect our results. Nevertheless, the clear and consistent trends we observe remain robust and physically meaningful. Our hierarchy of secondary bias factors holds despite these limitations, and our findings provide valuable observational benchmarks for testing theoretical models of halo formation.

\begin{figure}
    \centering
    \includegraphics[width=1\columnwidth]{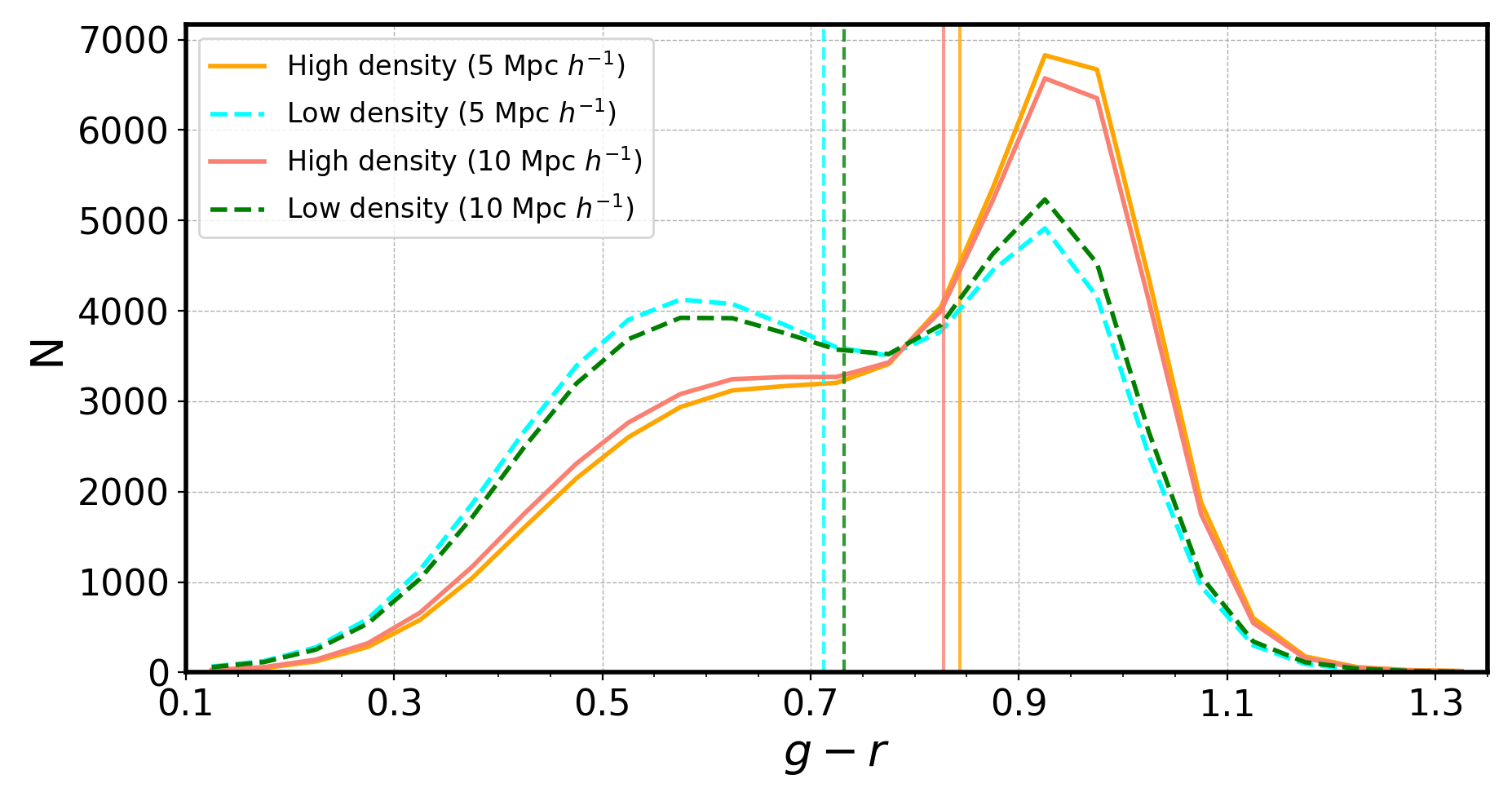}	

    \caption{Colour distribution of central galaxies with different density environments. The vertical lines indicate the medians of the respective samples. }
    \label{fig:summary2}
\end{figure}

The results presented in this paper provide a robust observational perspective on environmental secondary bias that, at face value, appears to be in slight tension with the conclusions of \cite{alam2019}, who found no evidence of a direct effect of the cosmic web beyond its modulation of the halo mass function. Methodological differences might, however, explain these apparent discrepancies. While \cite{alam2019} analysed individual galaxies and inferred halo mass using a halo occupation distribution model, our direct use of a cluster catalogue allows for more precise control of halo mass, which enables us to disentangle and more clearly detect the direct influence of cosmic web geometry on the clustering of fixed-mass haloes. This is consistent with \cite{Izzo2025}, who found minimal HOD variation across environments, confirming that the cosmic web influences halo clustering while only minimally altering their internal galaxy content. Our findings complement these previous results by demonstrating that a secondary bias signal emerges once we effectively control for halo mass. 

To explore the connection between the observed environmental secondary bias signals and potential assembly bias effects, we analysed the relationship between our selected subsamples and galaxy colour, a property expected to correlate with halo assembly history \citep{Hearin2013, Wechsler2018, Behroozi2019, ChavesMontero2020, MonteroDorta2021}. Figure~\ref{fig:summary2} presents this analysis, showing the colour distributions of galaxies across different density environments. The clear separation in colour between high- and low-density subsets suggests that environmental secondary bias based on local density may be linked to halo assembly bias through halo accretion history. These results align with the broader picture in which (local) environmental diagnostics act as mediators between internal halo properties and large-scale bias, as indicated by some hydrodynamical simulation results \citep{Paranjape2018, MonteroRodriguez2024}.

However, as shown in Fig.~\ref{Dist-density-color}, we do not find a clear connection between the distance of central galaxies to the critical points of the cosmic web and their colour, which complicates efforts to probe the aforementioned link based on our current cosmic web characterisation. Although simulations show that the relation between colour (or assembly history) and DisPerSE-derived distances is weak, \citet{MonteroRodriguez2024} demonstrated that the halo assembly bias signal changes significantly when the sample is conditioned on proximity to critical points. Further investigation, incorporating additional cosmic web identifiers, will be necessary to clarify whether these results are affected by the specific characteristics of the adopted cosmic web descriptor. 

In the future, it would be interesting to investigate how two-halo galactic conformity relates to the secondary bias trends reported here, especially those involving central galaxy colour and environment \citep [e.g.,][]{Kauffmann2015, Hearin2016,Hatfield2017, Lacerna2022, Lacerna2025}.

In conclusion, our study provides comprehensive observational evidence that secondary bias is a multi-faceted and widespread phenomenon in the galaxy distribution. We have established a definitive hierarchy of secondary bias drivers: local environmental measures, particularly overdensity, produce the strongest signals, followed by intrinsic galaxy properties like color, with cosmic web geometry generating more subtle though significant effects. This hierarchy underscores the complex interplay between local environmental processes and larger-scale cosmic web structure in shaping halo clustering. The consistent detection of these signals across multiple environmental metrics confirms that a halo's location within the cosmic web is a fundamental determinant of its clustering strength, independent of mass. Furthermore, our identification of color bias and its mediation through local density provides a direct observational link to galaxy assembly bias, demonstrating that secondary dependencies at fixed halo mass are not merely theoretical predictions but observable features of the galaxy population. Our work thus contributes to the growing body of observational evidence for assembly bias, joining findings from anisotropic clustering \citep{Obuljen2019, Obuljen2020}, halo occupancy variations \citep{Yuan2021, Wang2022, Pearl2024}, the stellar-to-halo mass relation \citep{Oyarzun2024}, and spin-dependent clustering \citep{Kim2025}. The robust quantification of these effects presented here provides crucial benchmarks for theoretical models and highlights the importance of accounting for secondary bias in precision cosmological analyses.

\begin{acknowledgements}
      We thank the anonymous referee for their constructive comments and suggestions, which helped to improve the clarity and quality of this manuscript.
      FR thanks the support by Agencia Nacional de Promoci\'on Cient\'ifica y Tecnol\'ogica, the Consejo Nacional de Investigaciones Cient\'{\i}ficas y T\'ecnicas (CONICET, Argentina) and the Secretar\'{\i}a de Ciencia y Tecnolog\'{\i}a de la Universidad Nacional de C\'ordoba (SeCyT-UNC, Argentina). ADMD acknowledges support from the Universidad Técnica Federico Santa María through the Proyecto Interno Regular \texttt{PI\_LIR\_25\_04}.
      FR and ADMD thank the ICTP for their hospitality and financial support through the Junior Associates Programme 2023–2028 and Regular Associates Programme 2022–2027, respectively.
\end{acknowledgements}


\bibliographystyle{aa} 
\bibliography{main} 
\end{document}